\documentclass[12pt]{article}
\usepackage[paperwidth=8.5in,paperheight=11in,left=1in,right=1in,top=1in,bottom=1in]{geometry}
\usepackage{amsmath,amssymb}
\usepackage{booktabs}
\usepackage{graphicx}
\usepackage{array}
\usepackage[hidelinks]{hyperref}
\usepackage{amsthm}
\usepackage{setspace}
\usepackage{tikz}\usetikzlibrary{positioning,arrows.meta,calc}
\theoremstyle{definition}
\newtheorem{remark}{Remark}

\newcommand{\ail}{{\rm ail}}
\newcommand{\rud}{{\rm rud}}
\newcommand{\ps}{{p_{\rm s}}}
\newcommand{\rs}{{r_{\rm s}}}

\title{\Large\bfseries Wing Rock Modeling and Control\\ for Flight-Critical Systems}
\author{Eugene Lavretsky\thanks{\raggedright Senior Principal Technical Fellow, Boeing Research \& Technology; AIAA Fellow; eugene.lavretsky@boeing.com.}\\
{\normalsize The Boeing Company, Huntington Beach, California 92647}\\[10pt]
Miroslav Krstic\thanks{\raggedright Distinguished Professor, Department of Mechanical and Aerospace Engineering; corresponding author, mkrstic@ucsd.edu.}\\
{\normalsize University of California San Diego, La Jolla, California 92093}}
\date{}

\begin{document}
\maketitle
\begin{abstract}
\noindent Aircraft equipped with delta wings are designed to be aerodynamically
efficient in high-speed flight. However, at low-speed flight conditions with
sufficiently high angle of attack, these vehicles often have lightly damped,
close to unstable lateral-directional dynamics, such as the dutch roll mode.
As a result, a delta wing aircraft often exhibits the so-called wing rock
(WR) phenomenon, with a back and forth rocking motion, while the aircraft
sideslip angle, the roll and the yaw angular rates remain small. In this
paper, the WR dynamics are defined as a lightly damped dutch roll mode
excited by airmass vortical external input, acting at or near the dutch
roll natural frequency. The vortical signal comes from Fourier-Galerkin
approximation of the Navier-Stokes equation, with no-penetration boundary
conditions and the resulting airmass velocities are calculated by the
Lorenz dynamical system of three coupled bilinear ordinary differential
equations. In that context, the Lorenz bilinear dynamics become the
disturbance-generator exosystem, acting on the lightly damped
lateral-directional aircraft mode at or near its natural frequency, and
thus exciting it to produce near oscillatory wing rocking motion. Using the
proposed model, a flight control design trade study is conducted to
mitigate and suppress the WR phenomenon.
\end{abstract}

\section{Introduction}

\noindent\textbf{Problem.} Wing rock (WR) was named before it was modeled.
The NASA reports on fighter aircraft at high angle of attack define it by
what is observed: ``slightly damped lateral oscillations about the
longitudinal body axis'' \cite{CG77}; ``an undamped Dutch roll mode above
$\alpha=17^\circ$ due to loss of $C_{l_p}$ and $C_{n_r}$, while
$C_{l_\beta}$ remains high. The oscillation is almost entirely about the
roll axis and the term `wing rock' is often used to describe this type of
motion'' \cite{NGO80}. The first analytical model retains the descriptive
definition, ``an uncommanded roll-yaw oscillation dominated by roll
motion'' \cite{HL85}, and the flight manual of the F-14 records that ``even
small control inputs will cause yaw oscillations that, if unchecked, can
produce a mild wing rock'' \cite{W92}. WR appears at the edge of the
maneuvering envelope, near $1g$ with the angle of attack held above
$20^\circ$, whereby a pilot and/or a flight control system opposes it with
lateral stick and rudder inputs \cite{NWB85,NOG79}. From the flight dynamics
perspective, the WR oscillations are undesirable and a flight critical
control system would be designed to suppress these unwanted oscillations. Such a control system must satisfy common
requirements on margins, bandwidth and attenuation of structural modes, all
well-defined and imposed by aircraft certification standards. A model-based
design and analysis of a flight critical control system for WR suppression
entirely depends on a WR model selected for the design.

\noindent\textbf{WR as a limit cycle.} Known models of WR treat it as a
nonlinear limit cycle oscillation of the aircraft roll degree of freedom:
at a sufficiently high angle of attack, the leading-edge vortices of a
slender delta wing separate asymmetrically and induce the WR motion
\cite{Ericsson84,HL85,EMN89,AN94}. The reviews
\cite{Ericsson95,Mabey97,Katz99,NP03,GGV05} cover the experiments, the
vortex mechanism and the models; the WR mechanism is re-examined for a flying wing in \cite{LFW23}. The models are of Van der Pol type, a roll equation with amplitude-dependent
damping, single-degree-of-freedom in the slender-wing studies, two degrees
of freedom in \cite{GR02}, with the flow lagged in the aerodynamic states
of Goman and Khrabrov \cite{GK94}. The same premise
appears as a Hopf bifurcation of the coupled equations, in the F-15
\cite{LD97}, the F-14 \cite{W92}, and in computed flow \cite{S00}. The
control designs in the literature use those models and, with a few
exceptions, the roll equation alone. Control design suppression methods for
WR include optimal control \cite{LL93}, adaptive and neural control
\cite{SYW95,LTWI22}, adaptive feedback linearization \cite{MK96},
backstepping with a disturbance observer (DO) on a lateral-directional model with
lagged aerodynamic states \cite{WCGW17}; post-stall control with unsteady
models \cite{LCZSQ18}, and lift regulation at the airfoil level
\cite{SJL20}.

\noindent\textbf{WR as a forced response.} Menner and Lavretsky
\cite{ML26} model the unsteady aerodynamics differently: not as a lag on
the motion of the aircraft but as a vortical flow in the airmass, the aircraft
induced airmass convection driven by the lift force acting back on the
airmass, according to Newton's third law of motion. It is shown that the
Lorenz system \cite{Saltzman62,Lorenz63} forced by the aircraft aerodynamic lift represents
the minimal order approximation/truncation of the Navier--Stokes (NS)
solution for the Rayleigh--B\'enard problem with no-penetration boundary
conditions. In addition, the related airmass dynamics become unsteady and
deterministically chaotic when the lift exceeds its critical value.
Based on that, we build a modeling paradigm for the WR phenomenon. At low
speeds and high angles of attack, aircraft lateral-directional dynamics are
benign, with sufficiently small deviations from a trimmed condition, and
because of that the lateral-directional system is essentially linear time
invariant (LTI). Also, at these conditions, the dutch roll is lightly
damped, and the Lorenz vortical flow, whose dominating frequency is near
the dutch-roll frequency, drives the dutch-roll mode at resonance. We use
no nonlinearity in the aircraft lateral-directional dynamics. Instead, we
postulate that WR is the forced response of a lightly damped dutch roll
mode to airmass induced vorticity disturbance. The Air Force classification of WR admits this
kind: ``small-amplitude intermittent roll oscillations \ldots\ Here the roll
motion is not periodic. This type of WR is normally associated with
low airspeed, high angle-of-attack (AOA) flight in gusty conditions''
\cite{Nolan92}. We also show that a controller can reduce the undesired
WR transients but not necessarily remove their effect, since the vortical
airmass sideslip angle adds to the vehicle sideslip angle almost instantaneously,
without a time-delay in between the two signals.

\begin{table}[!t]
\centering
\caption{Key notation.}
\label{tab:notation}
\renewcommand{\arraystretch}{1.15}
\begin{tabular}{p{3.3cm}p{11.2cm}}
\hline\hline
\textbf{Symbol} & \textbf{Meaning}\\
\hline
$A$, $B$ & aircraft state and input matrices, (\ref{eq:AB})\\
$D$ & input gain matrix of the normal form, (\ref{eq:Dmat})\\
$F_0=\rho gN_z$ & forcing of the Lorenz block, slug-ft units\\
$G$, $H(s)$ & gain and filter of the compensator (\ref{eq:comp})\\
$K$; $c_1,c_2,c_3$ & LQ gain; backstepping gains\\
$L_{(\cdot)}$, $N_{(\cdot)}$, $Y_{(\cdot)}$ & roll moment, yaw moment and side force derivatives, s$^{-2}$, s$^{-1}$, ft/s$^2$\\
$N_y$; $N_z$ & lateral load factor, (\ref{eq:Ny}); normal load factor\\
$V_T$ & airspeed, ft/s\\
$X$, $Y$, $Z$ & Lorenz states of the airmass convection\\
$d=\dot\beta_v$, $\hat d$ & vortex rate and its DO estimate, s$^{-1}$\\
$k$ & DO bandwidth, s$^{-1}$\\
$\ps$, $\rs$; $p_b$ & stability-axis roll and yaw rates; body-axis roll rate, s$^{-1}$\\
$u=(\delta_\ail,\delta_\rud)$ & control input\\
$x=(\beta,\ps,\rs)$, $x_p$ & aircraft state; the same with inertial sideslip $\beta_b$\\
$y_r$, $y_{\rm c}$ & regulated output (\ref{eq:yr}) and its command\\
$y$, $z$ & lateral and vertical position of the airmass origin in the body frame, ft\\
$\alpha$ & angle of attack\\
$\beta$, $\beta_b$, $\beta_v$ & sideslip (airmass-relative), inertial sideslip, vortex sideslip angles\\
$\gamma_y$, $\gamma_z$ & cell velocity amplitudes per unit $X$, ft/s\\
$\delta_\ail$, $\delta_\rud$ & aileron and rudder deflections\\
$\zeta_1$, $\zeta_2$, $\eta$ & normal-form coordinates, (\ref{eq:coord})\\
$\zeta_u$ & side force of the surfaces in (\ref{eq:beta})\\
$\zeta$, $\omega_n$ & dutch-roll damping ratio and natural frequency, s$^{-1}$\\
$\theta_{ij}$ & Lorenz block constants, s$^{-1}$\\
$\omega_y$, $\omega_z$ & inverse width and height of the convection cell, ft$^{-1}$\\
\hline\hline
\end{tabular}
\end{table}

\noindent\textbf{Contributions.}
\begin{itemize}\itemsep2pt
\item We model WR as a forced response of the dutch-roll mode with three
components: a linear lateral-directional aircraft dynamics, the bilinear
Lorenz system of convection coefficients forced by the aircraft lift force
per unit volume, and the bilinear motion of
the airmass origin in the velocity field of the convection cell, which
converts the convection into a persistent vortex induced sideslip
increment when the lift force exceeds its convection-critical value. The Lorenz dominant frequency falls near the dutch-roll
frequency, and on the F-16 at $\alpha=30^\circ$, with dynamic pressure at
$50$~psf, the model produces the WR oscillations observed in the aircraft flight
records.
\item We design a controller for lateral-directional dynamics in which
the vortex induced sideslip disturbance is unmatched: the vortex sideslip
enters the sideslip equation, where effectiveness of the aircraft control
surfaces (aileron and rudder) is minimal. We combine linear feedback with a
DO of order one, built from the sideslip equation alone, whose estimate is
routed to the surfaces through a static feedforward gain, and so the DO
contribution is absent from the loop gain at the surfaces and the resulting
stability margins are those of the linear feedback for every DO gain. A
trade study with actuators in the loop supports the design. It yields
$6$--$8$~dB loop gain attenuation at the dutch roll frequency.
\item We test the hypothesis if a DO of order one is ``too simplistic'' to
mitigate a chaotic vortex sideslip disturbance, by working out the exact
alternative: a nonlinear output regulation with an internal model of the
Lorenz-induced chaos. We show that such a design is possible when the
Lorenz airmass parameters are known. The design includes a closure of the
chain of time derivatives of the vortex rate or it adds a
Kazantzis--Kravaris/Luenberger filter of order six to ten. It turns out
that the resulting theoretical injection gains are too large to be
practical and would create an extreme sensitivity to measurement noise.
Since the baseline DO design already suppresses WR by $95\%$, we do not
conduct tests of a $100\%$ rejection with the output regulator and only
present it to inform the reader of the cost of
aiming for a perfect rejection of the vortex.
\end{itemize}

\noindent\textbf{Organization.} Section~2 defines the WR model, tests it
open-loop, and explains its structure. In Section~3, several controllers
for WR suppression are introduced: the DO, its use with backstepping, with
LQ and with the LQR PI servo. These systems are analyzed to ensure that
adequate tracking performance and stability margins are attained.
Section~4 is the trade study to demonstrate attenuation by LQ+DO against LQ
alone. Section~5 works out the exact rejection of the vortex sideslip
disturbance by output regulation and what it
adds over the DO; Section~6 concludes. The appendix contains supporting computations: flight conditions,
aircraft data, the airmass origin, and the model closure order.

\section{Model and control structure}

\subsection{Model}

The notation is listed in Table~\ref{tab:notation}. The aircraft model represents the lateral-directional aircraft dynamics,
(\ref{eq:beta})--(\ref{eq:r}),
typically used for inner-loop flight control design, with state
$x=(\beta,\ps,\rs)$, where $\beta$ is the aircraft true sideslip angle relative to the incoming airmass,
$\ps$, $\rs$ are the stability-axis roll and yaw rates, $p_b$ is the roll
rate in body axes, and the control
input $u=(\delta_\ail,\delta_\rud)$ is comprised of aileron and rudder
surface deflections, respectively \cite{LW}. The lateral force and
roll/yaw moments depend on the true
sideslip $\beta=\beta_b-\beta_v$, with $\beta_b$ the inertial sideslip and
$\beta_v$ the sideslip increment due to the vortical flow of the airmass,
caused by the aircraft and not by the environment, so the vortex sideslip
enters the sideslip equation as its rate increment:
\begin{subequations}\label{eq:model}
\begin{eqnarray}
\dot\beta &=& \dfrac{Y_\beta}{V_T}\beta+\dfrac{Y_{\ps}}{V_T}\ps
              +\Big(\dfrac{Y_{\rs}}{V_T}-1\Big)\rs
              +\dfrac{Y_{\delta_\ail}}{V_T}\delta_\ail
              +\dfrac{Y_{\delta_\rud}}{V_T}\delta_\rud-\dot\beta_v \label{eq:beta}\\
\dot p_{\rm s} &=& L_\beta\beta+L_{\ps}\ps+L_{\rs}\rs
              +L_{\delta_\ail}\delta_\ail+L_{\delta_\rud}\delta_\rud \label{eq:p}\\
\dot r_{\rm s} &=& N_\beta\beta+N_{\ps}\ps+N_{\rs}\rs
              +N_{\delta_\ail}\delta_\ail+N_{\delta_\rud}\delta_\rud \label{eq:r}\\
\dot X    &=& -\theta_{11}X+\theta_{12}Y \label{eq:X}\\
\dot Y    &=& \rho g N_z(t)\,\theta_{21}\,X-\theta_{22}Y-\theta_{23}XZ \label{eq:Y}\\
\dot Z    &=& -\theta_{31}Z+\theta_{32}XY \label{eq:Z}\\
\dot y    &=& -\gamma_yX\sin(\pi\omega_yy)\cos(\pi\omega_zz)-p_b\,z \label{eq:y}\\
\dot z    &=& \gamma_zX\cos(\pi\omega_yy)\sin(\pi\omega_zz)+p_b\,y \label{eq:z}\\
\beta_v   &\approx& -\dfrac{\gamma_y}{V_T}\,X\sin(\pi\omega_yy)\cos(\pi\omega_zz) \label{eq:bg}\\
p_b       &=& \ps\cos\alpha-\rs\sin\alpha \label{eq:pb}
\end{eqnarray}
\end{subequations}
Equations (\ref{eq:beta})--(\ref{eq:r}) represent the aircraft
lateral-directional dynamics driven by aileron and rudder surfaces, while
excited by the vortex induced sideslip rate.
Equations (\ref{eq:X})--(\ref{eq:Z}) are the Lorenz dynamics of the airmass
convection, the three-mode Fourier--Galerkin truncation of the incompressible
flow equations between two boundaries \cite{Saltzman62,Lorenz63}, derived
for the airmass over the wing in \cite{ML26}: $X$ is the amplitude of the
stream function of the circulation, which sets the airmass velocity, and
$Y$, $Z$ are the amplitudes of the first and the second harmonic deviation
of the lift-induced forcing function from its nominal distribution. Equations
(\ref{eq:y})--(\ref{eq:z}) are the airmass dynamics: the position $(y,z)$ of
the airmass origin relative to the body-fixed frame at the center of
gravity, $z$ negative above the wing, moved by the velocity field of the
convection cell, lateral $v_v=-\gamma_yX\sin(\pi\omega_yy)\cos(\pi\omega_zz)$
and vertical $w_v=\gamma_zX\cos(\pi\omega_yy)\sin(\pi\omega_zz)$, plus the
rotating-frame term $\omega_b\times(0,y,z)$ of the rotating body frame, with
$p_b$ from (\ref{eq:pb}), $\alpha$ the aircraft angle of attack, and $N_z$
the vehicle vertical load factor; through $p_b$ the airmass block is directly
driven by the aircraft. The airmass cell has width $1/\omega_y$ and height $1/\omega_z$, and its velocity field
is incompressible when $\gamma_y\omega_y=\gamma_z\omega_z$. Equation
(\ref{eq:bg}) is the vortex sideslip $\beta_v$, the angle by which the vorticial lateral velocity
component $v_v$ tilts the oncoming air, while the vertical vorticial speed
$w_v$ acts on the longitudinal axis and is not used
here.

On a tactical aircraft, the true sideslip angle often is not measured.
Instead, it is estimated using the vehicle lateral acceleration, angular
rates and control surfaces. However, for the sake of clarity and as a
simplification, in this paper we assume that $\beta$ is a known
measurement. In compact form, (\ref{eq:beta})--(\ref{eq:r}), without the vortex
sideslip rate, is a linear time-invariant (LTI) system of the form
\begin{equation}
\dot x=Ax+Bu,
\label{eq:lti}
\end{equation}
with the vehicle stability and control (S\&C) derivative data computed at
the selected flight condition,
\begin{equation}
A=\begin{pmatrix}
\dfrac{Y_\beta}{V_T}&\dfrac{Y_{\ps}}{V_T}&\dfrac{Y_{\rs}}{V_T}-1\\[4pt]
L_\beta&L_{\ps}&L_{\rs}\\
N_\beta&N_{\ps}&N_{\rs}
\end{pmatrix},\qquad
B=\begin{pmatrix}
\dfrac{Y_{\delta_\ail}}{V_T}&\dfrac{Y_{\delta_\rud}}{V_T}\\[4pt]
L_{\delta_\ail}&L_{\delta_\rud}\\
N_{\delta_\ail}&N_{\delta_\rud}
\end{pmatrix},
\label{eq:AB}
\end{equation}

\noindent\textbf{Vertical load factor.} $F_0=\rho gN_z$ in (\ref{eq:Y}) is the
forcing input signal of the Lorenz block, in the derivation of \cite{ML26}, the lift
reaction on the airmass, where $N_z$ is the aircraft vertical load factor, in g. $N_z$ is a longitudinal quantity,
set by the aircraft angle of attack and pitch rate, which do not appear in
(\ref{eq:beta})--(\ref{eq:r}); no state of this model appears in $F_0$, and it is taken as a
given signal, with $N_z=1$ in the simulations. The control designs of Section~3 do not use the airmass
model and are unaffected by $N_z(t)$ or by the terms in $p_b$; Section~\ref{sec:or} takes
$N_z$ as a constant and $p_b=0$, since with either the exosystem becomes
non-autonomous with a known input.
Adding the pitch states $(\alpha,q)$ with the elevator and $N_z$ as their
output, from Chapter~1 of \cite{LW}, is the extension that would close this
loop.

\noindent\textbf{Linearized vortex angle.} In practical applications,
sideslip deviations during WR are small \cite{NGO80,HL85}, and therefore the
arcsin function in $\beta_v=\arcsin(v_v/V_T)$ can be dropped, as in
(\ref{eq:bg}).

\noindent\textbf{Lateral load factor.} The lateral load factor, in g, is
\begin{equation}
N_y=\frac{1}{g}\big(Y_\beta\beta+Y_{\ps}\ps+Y_{\rs}\rs
+Y_{\delta_\ail}\delta_\ail+Y_{\delta_\rud}\delta_\rud\big).
\label{eq:Ny}
\end{equation}

\noindent\textbf{Control side forces.} $Y_{\delta_\ail}$,
$Y_{\delta_\rud}$ are sufficiently small and dropped in the designs below,
except for the servo-controller of Section~\ref{sec:servo}; they are retained in that servo
and in the plant simulated in Section~\ref{sec:trackff}, and the DO of
Section~\ref{sec:do} contains them through the measured $N_y$.

\noindent\textbf{Parameters and states.} The S\&C parameters of
(\ref{eq:beta})--(\ref{eq:r}), $V_T$, $\rho$ and $g$ are known. The ten
airmass constants $\theta_{ij}$, $\gamma_y$, $\omega_y$, $\omega_z$ are
known and positive, with $\gamma_z=\gamma_y\omega_y/\omega_z$. The
states $X$, $Y$, $Z$, $y$, $z$ are not available as system output
measurements, while $\beta$, $\ps$, $\rs$, $N_y$, $N_z$ and $\alpha$ are measured.

\subsection{WR as a forced response of the aircraft dutch roll mode}\label{sec:forced}

All simulation tests in Sections~\ref{sec:forced}--\ref{sec:or}, and the margins of
Section~\ref{sec:margins}, are defined under one of the two conditions of
Table~\ref{tab:cond}.

\begin{table}[!t]
\caption{Simulation test conditions: Condition~I, the flight condition with the identified cell; Condition~II, the persistent vortex with the nondimensional Lorenz block.}
\label{tab:cond}
\centering
\renewcommand{\arraystretch}{1.12}
\begin{tabular}{>{\raggedright\arraybackslash}p{4.1cm}>{\raggedright\arraybackslash}p{5.3cm}>{\raggedright\arraybackslash}p{5.3cm}}
\toprule
& Condition~I & Condition~II\\
\midrule
used in & Sections~\ref{sec:airmass}--\ref{sec:track} & Sections~\ref{sec:margins}, \ref{sec:persist}, \ref{sec:or}\\
aircraft & F-16 tables, $\alpha=30^\circ$, $50$~psf, $V_T=205$~ft/s, (\ref{eq:highalpha}) & F-16, $\alpha=30^\circ$, $V_T=230$~ft/s, first set of Appendix~\ref{app:data}\\
dutch roll $\omega_n$, $\zeta$ & $2.09$~s$^{-1}$, $0.25$; variant $0.08$ & $1.12$~s$^{-1}$, $0.76$\\
airmass & identified cell, $2.8\times2.0$~ft & Lorenz constants $\times\,0.1$~s$^{-1}$, one-dimensional origin\\
vortex rate $\omega_L$ & $2.4$~s$^{-1}$ & $1.16$~s$^{-1}$\\
vortex sideslip $\beta_v$, rms & $0.40^\circ$ & $2.33^\circ$\\
DO bandwidth $k$ & $20$~s$^{-1}$ & $20$~s$^{-1}$\\
actuators & second order, $60$~rad/s, Sections~\ref{sec:sweep}--\ref{sec:track}; rate and position limits in Section~\ref{sec:track} & none\\
\bottomrule
\end{tabular}
\end{table}

\subsubsection{Airmass dynamics: step test and persistence of the vortex sideslip}\label{sec:airmass}

The airmass parameters, identified for the flight condition of this
section, are $\theta_{11}=0.9771$,
$\theta_{12}=276.86$, $\theta_{21}=0.55921$, $\theta_{22}=0.16064$,
$\theta_{23}=3.5136$, $\theta_{31}=0.42837$, $\theta_{32}=7.0273$, in
s$^{-1}$; $\gamma_y=6.3049$, $\gamma_z=4.4583$~ft/s, $\omega_y=0.35478$,
$\omega_z=0.50173$~ft$^{-1}$, a cell $2.8$~ft wide and $2.0$~ft high, with
$\gamma_y\omega_y=\gamma_z\omega_z$; $F_0=\rho gN_z$ with
$F_0^{\rm crit}=0.03$, $\rho g=0.0765$ at sea level in slug--ft units, so that
$F_0=0.0765N_z$ is above the critical value at $N_z=1$; units are English,
angles in radians. The origin of the Lorenz model is unstable at
$F_0>\theta_{11}\theta_{22}/(\theta_{12}\theta_{21})=0.00101$;
$F_0^{\rm crit}$ is $30$ times that, the classic ratio at which the Lorenz
attractor appears, and $F_0=0.0765$ is $75$ times it. At the nonzero
equilibrium for $F_0=0.0765$ the Jacobian has eigenvalues $-1.71$ and
$0.07\pm2.42j$: the vortex rate is about $2.4$~s$^{-1}$, twice that of Condition~II.

\begin{figure}[p]
\centering
\includegraphics[width=0.86\textwidth]{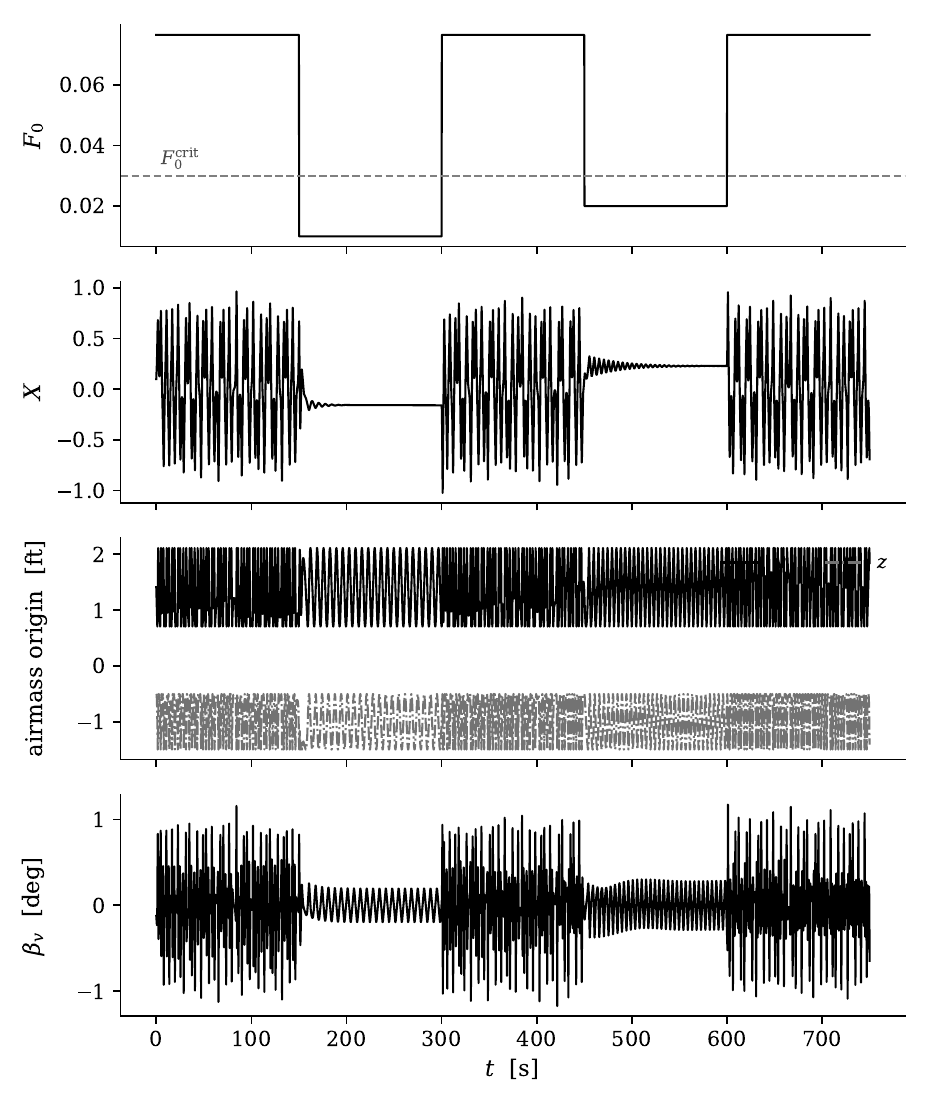}
\caption{Airmass dynamics, $p_b=0$, origin started at $(1.41,-0.5)$~ft, $F_0$
stepped across $F_0^{\rm crit}=0.03$: $X$ chaotic above, constant below; the
vortex sideslip persists.}
\label{fig:exo}
\end{figure}

\begin{figure}[t]
\centering
\includegraphics[width=0.8\textwidth]{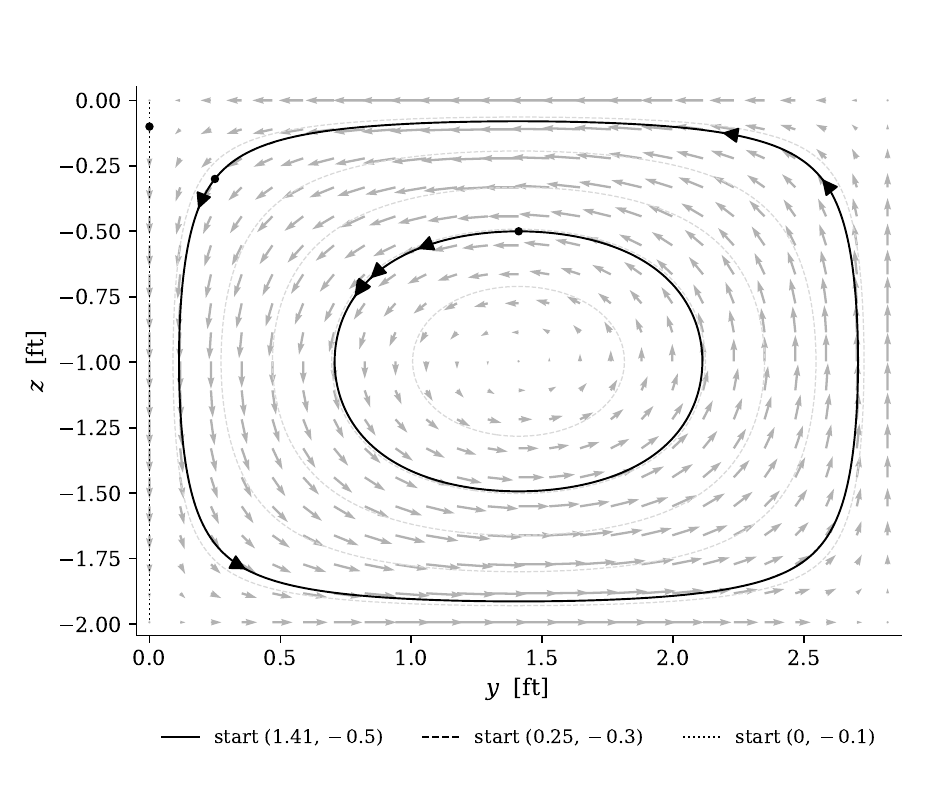}
\caption{Airmass origin in the cell, $p_b=0$: velocity field for $X>0$,
streamlines, and sixty seconds of trajectory from three starts, one on the
boundary.}
\label{fig:phase}
\end{figure}

Figure~\ref{fig:exo} shows the airmass dynamics, with $p_b=0$, the system
trajectory starting inside the cell at $(y,z)=(1.41,-0.5)$~ft, and $F_0$ step-changing
$0.0765\to0.010\to0.0765\to0.020\to0.0765$ at $150$~s intervals. $X$ is chaotic above
$F_0^{\rm crit}$, settles to a constant nonzero value below it, and is
chaotic again when $F_0$ returns; at $F_0=0.020$ the settling is preceded by a long chaotic
transient, as for the Lorenz system between the homoclinic and the attractor
thresholds. The trajectory circulates on the closed streamline through its
start, $y\in[0.71,2.11]$, $z\in[-1.49,-0.50]$~ft (Figure~\ref{fig:phase},
Appendix~\ref{app:origin}), and the vortex sideslip
persists: $0.35^\circ$ rms, $1.2^\circ$ peak, in the chaotic intervals;
$0.13^\circ$ and $0.19^\circ$ rms, periodic, in the subcritical ones, where
the constant $X$ drives a steady circulation. On the cell boundary the vortex sideslip
is zero: from a start on it such as $(0,-0.1)$, $y\equiv0$ and
$\beta_v\equiv0$ until the aircraft rolls (Figure~\ref{fig:phase}, Appendix~\ref{app:origin}), so
we start the tests below inside the cell.

\begin{figure}[t]
\centering
\includegraphics[width=0.86\textwidth,trim=0 275 0 0,clip]{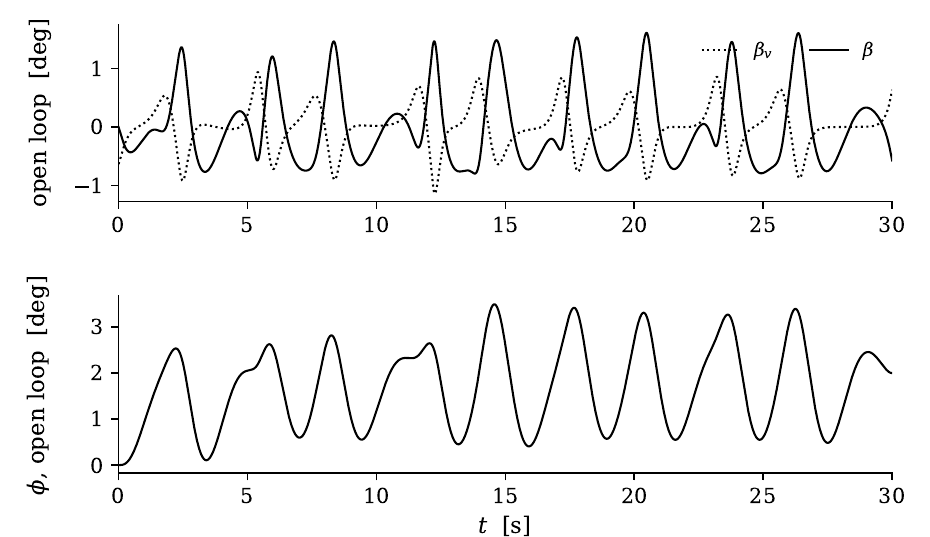}
\caption{F-16 at $\alpha=30^\circ$, $50$~psf, cell vortex, open loop from
rest: vortex sideslip and sideslip; bank angle.}
\label{fig:ol}
\end{figure}
\begin{figure}[!t]
\centering
\includegraphics[width=0.86\textwidth]{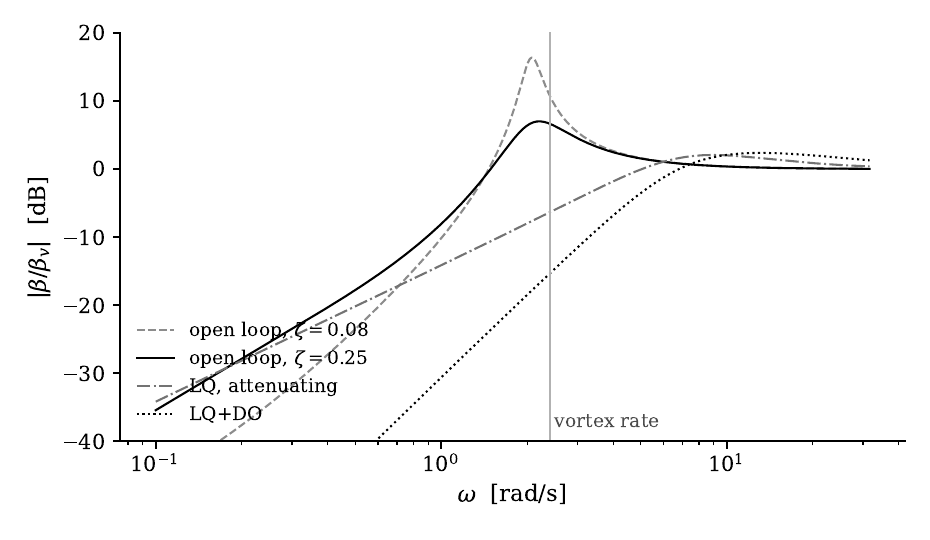}
\caption{Transfer from vortex angle to sideslip, $\alpha=30^\circ$, $50$~psf,
$Q_\beta=(1^\circ)^{-2}$: open loop at two dampings, LQ attenuation, LQ+DO;
vertical line, vortex rate.}
\label{fig:freq}
\end{figure}

\subsubsection{Open loop: resonance and forced response}\label{sec:openloop}

The paradigm under test is that the lateral-directional dynamics are linear for
sufficiently small sideslip angles such as $|\beta|<10^\circ$, the dutch roll at WR conditions is lightly damped,
and the Lorenz vortex, whose natural frequency with the parameters of Section~\ref{sec:airmass} is near the
dutch-roll frequency, drives it at resonance; whatever the
controller, the transfer from vortex sideslip to sideslip has a peak, and the oscillation
that remains there is the WR. The flight condition is
Condition~I of Table~\ref{tab:cond}, the aircraft from the NASA F-16
tables, Appendix~\ref{app:data}: dutch roll $-0.52\pm2.02j$,
$\omega_n=2.09$~s$^{-1}$, $\zeta=0.25$, roll $-0.15$. The tables do not give a near-neutral dutch roll at this condition; the
variant with the rotary damping derivatives reduced to one fifth,
$\zeta=0.08$ at the same $\omega_n$, is run alongside.

Figure~\ref{fig:freq} shows the transfer from $\beta_v$ to $\beta$, which is $s$
times the transfer from $d=\dot\beta_v$ to $\beta$. In open loop it
peaks at $\omega_n$, $6.9$~dB at $\zeta=0.25$ and $16$~dB at $\zeta=0.08$:
the resonance of the paradigm. Figure~\ref{fig:ol} is the time
domain: in open loop from rest over thirty seconds, with the bank angle
$\dot\phi=\ps/\cos\alpha$ as an output, the vortex sideslip is $0.40^\circ$ rms,
$1.1^\circ$ peak, and sideslip follows it with the amplification of the
resonance, $0.67^\circ$ rms, $1.6^\circ$ peak ($1.4^\circ$ rms, $3.0^\circ$
peak at $\zeta=0.08$), bank angle within $\pm3.5^\circ$ ($\pm6^\circ$). This
is a small oscillation at the dutch-roll period, not WR. From the
boundary start $(0,-0.1)$ there is no vortex sideslip at all, $y\equiv0$
(Appendix~\ref{app:origin}).

\begin{figure}[t]
\centering
\includegraphics[width=0.86\textwidth]{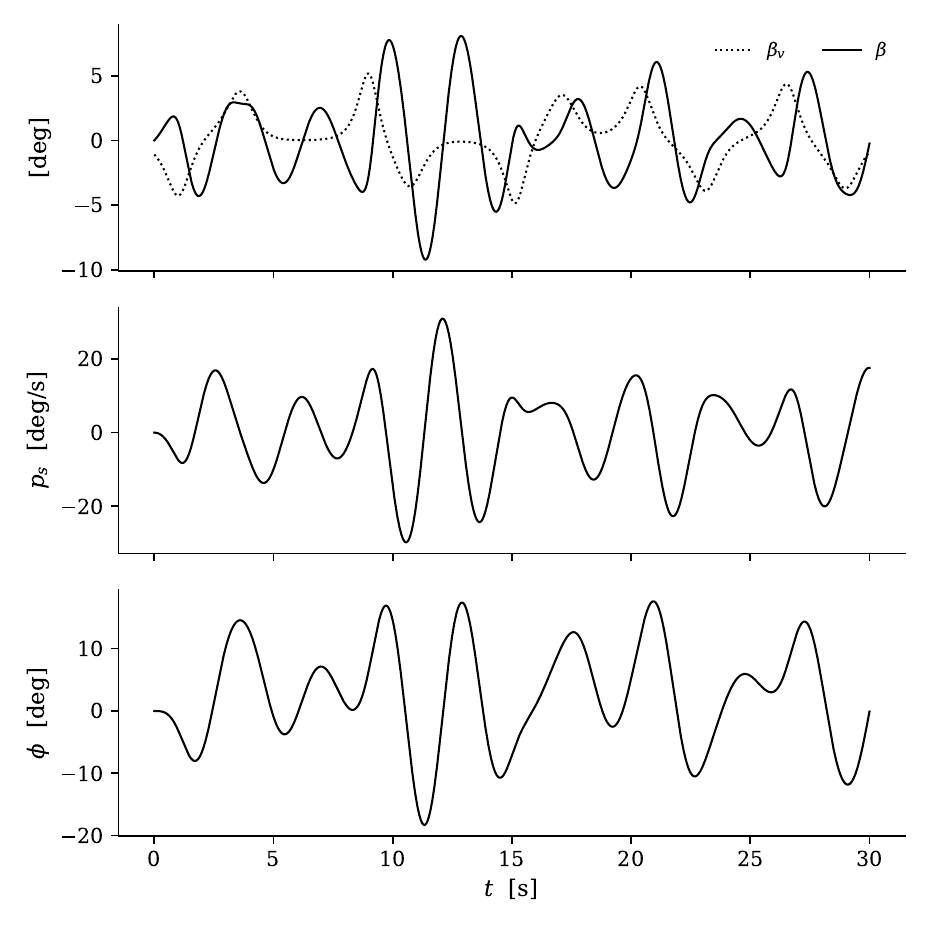}
\caption{Open loop, vortex sideslip scaled by three with position dependence removed,
$\zeta=0.08$: sideslip $\pm9^\circ$, roll rate $\pm30^\circ$/s, bank
$\pm18^\circ$.}
\label{fig:ol2}
\end{figure}

The WR amplitudes of the flight records need a vortex sideslip three times
larger. With $\beta_v=-3\gamma_yX/V_T$, the Lorenz output driving the aircraft
directly at its own rate $2.4$~s$^{-1}$, on the dutch roll at $\zeta=0.08$,
the response (Figure~\ref{fig:ol2}) is sideslip $\pm9^\circ$, roll rate
$\pm30^\circ$/s and bank $\pm18^\circ$ at a period of about three seconds, the amplitudes and
period of the F-14 flight records. That is the forced response of the
paradigm, and it is what the controller of Section~3, tested in
Section~\ref{sec:trade}, acts on; the cell as parametrized gives a third of it.

\subsection{Normal form}

With $\beta_v=0$, (\ref{eq:beta})--(\ref{eq:r}) is $\dot x=Ax+Bu$ with
(\ref{eq:AB}). With the control side forces dropped, the pair $(A,B)$ is controllable if
and only if
\begin{equation}
L_{\delta_\ail}N_{\delta_\rud}-N_{\delta_\ail}L_{\delta_\rud}\neq0,\qquad
\dfrac{Y_{\ps}}{V_T}L_{\delta}+\Big(\dfrac{Y_{\rs}}{V_T}-1\Big)N_{\delta}\neq0
\ \text{ for }\delta=\ail\text{ or }\rud
\label{eq:cond}
\end{equation}
The first condition is that the roll--yaw
control effectiveness determinant is nonzero; the second holds whenever one
surface produces yawing moment, since $Y_{\rs}/V_T\ll1$. Every aircraft
satisfies both throughout the envelope. Controllability from a single
surface is the cubic condition
$\det[\,b_\delta\ Ab_\delta\ A^2b_\delta\,]\neq0$, which has no simple
factorization. The surface must affect sideslip, which enters (\ref{eq:AB})
only through yaw rate: the rudder does so through $N_{\delta_\rud}$, the
aileron only through its adverse yaw $N_{\delta_\ail}$ and through
$N_{\ps}$. For the F-16 of Appendix~\ref{app:data} the determinant is $4.4$
with condition number $8$ for the rudder, and $0.43$ with condition number
$81$ for the aileron. We use both surfaces from this point forward.

The regulated output is sideslip and roll rate,
\begin{equation}
y=(\beta,\ \ps).
\label{eq:out}
\end{equation}
With the vortex rate
\begin{equation}
d=\dot\beta_v,
\label{eq:d}
\end{equation}
the coordinates
\begin{subequations}\label{eq:coord}
\begin{eqnarray}
\zeta_1&=&\beta \label{eq:z1}\\
\zeta_2&=&\dfrac{Y_\beta}{V_T}\beta+\dfrac{Y_{\ps}}{V_T}\ps
          +\Big(\dfrac{Y_{\rs}}{V_T}-1\Big)\rs \label{eq:z2}\\
\eta&=&\ps \label{eq:eta}
\end{eqnarray}
\end{subequations}
bring (\ref{eq:beta})--(\ref{eq:r}) to the normal form
\begin{subequations}\label{eq:nf}
\begin{eqnarray}
\dot\zeta_1&=&\zeta_2-d \label{eq:nf1}\\
\dot\zeta_2&=&a_\zeta^{\!\top}x+\mu_\ail\delta_\ail+\mu_\rud\delta_\rud
              -\dfrac{Y_\beta}{V_T}\,d \label{eq:nf2}\\
\dot\eta&=&a_\eta^{\!\top}x+L_{\delta_\ail}\delta_\ail+L_{\delta_\rud}\delta_\rud ,
            \label{eq:nf3}
\end{eqnarray}
\end{subequations}
in which $a_\zeta$ and $a_\eta$ are the rows of $A$ carried into the second
and third equations and the input gains are
\begin{subequations}\label{eq:D}
\begin{eqnarray}
\mu_\delta&=&L_{\delta}\dfrac{Y_{\ps}}{V_T}
              +N_{\delta}\Big(\dfrac{Y_{\rs}}{V_T}-1\Big) \label{eq:mu}\\
D&=&\begin{pmatrix}\mu_\ail&\mu_\rud\\ L_{\delta_\ail}&L_{\delta_\rud}\end{pmatrix},
\qquad
\det D=\Big(\dfrac{Y_{\rs}}{V_T}-1\Big)
          \big(L_{\delta_\ail}N_{\delta_\rud}-L_{\delta_\rud}N_{\delta_\ail}\big).
          \label{eq:Dmat}
\end{eqnarray}
\end{subequations}
From $(\delta_\ail,\delta_\rud)$ to (\ref{eq:out}) the vector relative
degree is $(2,1)$, whose sum, three, is the dimension of (\ref{eq:AB}), and
$D$ is nonsingular by (\ref{eq:cond}). Hence there are no zero dynamics,
and the pair (\ref{eq:AB}), (\ref{eq:out}) has no transmission zero. The disturbance is unmatched with
mismatch index one: the inputs act in (\ref{eq:nf2}) and (\ref{eq:nf3}),
the vortex rate in (\ref{eq:nf1}), one integrator upstream of them. The
matched part $-\frac{Y_\beta}{V_T}d$ in (\ref{eq:nf2}) is negligible
against the unmatched $-d$ in (\ref{eq:nf1}).

\section{Design: disturbance observer, backstepping, and LQR}

The DO of Section~\ref{sec:do} is employed both in the compact backstepping
design of Section~\ref{sec:bdo} and in the more general LQR design of
Section~\ref{sec:lqdo}.

\subsection{Disturbance observer (DO)}\label{sec:do}

In (\ref{eq:nf1}), the vortex rate $d$ is an unmeasured disturbance. With
the control side forces of (\ref{eq:beta}) retained,
$\dot\zeta_1=\zeta_2+\zeta_u-d$, where
$\zeta_u=\frac{Y_{\delta_\ail}}{V_T}\delta_\ail+\frac{Y_{\delta_\rud}}{V_T}\delta_\rud$
is the side force of the surfaces, and $\zeta_2+\zeta_u=\frac{g}{V_T}N_y-\rs$
with $N_y$ of (\ref{eq:Ny}), so $d=\frac{g}{V_T}N_y-\rs-\dot\zeta_1$. It
cannot be obtained from this formula directly since, while the
airmass-relative sideslip $\zeta_1=\beta$ is measured, its derivative
$\dot\zeta_1=\dot\beta$ is unavailable for measurement. We apply a lowpass filter to
compute $d=\frac{g}{V_T}N_y-\rs-\dot\zeta_1$ and treat
\begin{equation}
\boxed{\;\hat d=\dfrac{k}{s+k}\Big(\dfrac{g}{V_T}N_y-\rs\Big)-\dfrac{ks}{s+k}\,\zeta_1\;}
\label{eq:dobtf}
\end{equation}
as the estimate of $d$. This lowpass-filtered exact $d$ is known in the
literature as a ``disturbance observer'' (DO). Its state space realization is
\begin{subequations}\label{eq:dob}
\begin{eqnarray}
\dot{\rm p}&=&-k\,{\rm p}+k\Big(\dfrac{g}{V_T}N_y-\rs\Big)+k^2\zeta_1,\qquad \hat d={\rm p}-k\,\zeta_1
\label{eq:dobstate}\\
\hat d&=&\dfrac{k}{s+k}\,d \label{eq:dobd}
\end{eqnarray}
\end{subequations}
where ${\rm p}$ is the state of the lowpass filter.
Both transfer functions in (\ref{eq:dobtf}) are proper. The second transfer function,
$\dfrac{s}{s/k+1}$, is an approximate differentiator, so no measured
variable is differentiated directly. The DO's inputs $\beta$, $\rs$ and
$N_y$ are measured. In the DO, $N_y$ is the incremental lateral load factor,
the measured lateral load factor minus $\cos\theta\sin\phi$, which need not
be constant nor computed at a trim condition. With these inputs
(\ref{eq:dobd}) holds exactly, with the side forces retained and for any
actuators; with $\zeta_2$ in place of $\frac{g}{V_T}N_y-\rs$,
$\hat d=\frac{k}{s+k}(d-\zeta_u)$. In the designs, where the side forces are
dropped, $\zeta_u=0$ and the two coincide.

\noindent\textbf{Guideline for choosing the DO bandwidth $k$.} Let $\omega_L$
be the rate of the vortical flow, from (\ref{eq:X})--(\ref{eq:Z}), and
$\omega_c$ the crossover frequency of the sideslip loop closed by the state
feedback alone. The guideline for the user is
\begin{equation}
k\ge3\,\omega_c\qquad\mbox{where}\quad \omega_c>\omega_L .
\label{eq:kguide}
\end{equation}
The second requirement is on the state feedback, not on the DO: a
disturbance at a rate above the crossover reaches sideslip with no
attenuation by any feedback, so the loop must be faster than the flow it is
to reject. The first requirement is on the DO: by (\ref{eq:dobd}) the
estimate is $d$ through a lowpass filter of bandwidth $k$, and it must be
accurate over the band in which the feedback acts, up to $\omega_c$. At
$k=3\omega_c$ the filter has $5\%$ amplitude error and $18^\circ$ phase lag
at $\omega_c$, and no error of consequence at $\omega_L$. A larger $k$ adds
nothing to the rejection while multiplying the sideslip measurement noise
by $k$ in the term $-ks/(s+k)\,\zeta_1$ of (\ref{eq:dobtf}).

Nothing in this section uses the Lorenz structure: $d$ is treated as an
unknown differentiable signal, and (\ref{eq:dobtf}) reconstructs it from the
plant model alone. Nor would the structure narrow the spectrum, since a
Lorenz output is broadband. Attenuation by loop gain alone requires a decade
of bandwidth for each decade of rejection; the DO instead cancels $d$ where
it acts, using only that it is differentiable. The structure enters only if
the chain of derivatives of $d$ is closed instead of truncated, which is
output regulation, Section~\ref{sec:or}.

\subsection{Backstepping with disturbance observer (BDO)}\label{sec:bdo}

With the tracking errors, the desired value of the virtual control, and the
DO error
\begin{subequations}\label{eq:bkst}
\begin{eqnarray}
e_1&=&\zeta_1=\beta \label{eq:e1}\\
\zeta_2^{\rm des}&=&\hat d-c_1e_1 \label{eq:z2des}\\
e_2&=&\zeta_2-\zeta_2^{\rm des} \label{eq:e2}\\
e&=&\hat d-d \label{eq:e}
\end{eqnarray}
\end{subequations}
$\zeta_2$ is the virtual control for (\ref{eq:nf1}), and the surfaces act one
step later, in (\ref{eq:nf2}); that step is the integrator between the
inputs and (\ref{eq:nf1}). The control law is
\begin{equation}
\boxed{\;\begin{pmatrix}\delta_\ail\\ \delta_\rud\end{pmatrix}
=D^{-1}\begin{pmatrix}
-c_1\big(\zeta_2-\hat d\big)-e_1-c_2e_2-a_\zeta^{\!\top}x
+\dfrac{Y_\beta}{V_T}\hat d\\[6pt]
-c_3\,\eta-a_\eta^{\!\top}x
\end{pmatrix}\;}
\label{eq:law}
\end{equation}
Eliminating $\dot{\rm p}$ through (\ref{eq:dobstate}) gives
$\dot{\rm p}=k(\zeta_2-\hat d)$, which is what has been used in
(\ref{eq:law}): the DO bandwidth $k$ cancels there and appears only in
(\ref{eq:dobtf}). The full-information version of (\ref{eq:law}) replaces
$\hat d$ by $d$ and adds $\dot d$ to the first row, under which sideslip is
rejected exactly. The closed loop in the error coordinates is
\begin{subequations}\label{eq:target}
\begin{eqnarray}
\dot e_1&=&-c_1e_1+e_2+e \label{eq:t1}\\
\dot e_2&=&-e_1-c_2e_2+\Big(k+c_1+\dfrac{Y_\beta}{V_T}\Big)e \label{eq:t2}\\
\dot \eta&=&-c_3\eta \label{eq:t3}\\
\dot e&=&-k\,e-\ddot\beta_v \label{eq:t4}
\end{eqnarray}
\end{subequations}
$\ps\to0$ exactly; $\beta$ is driven by the DO error $e$ through
(\ref{eq:t1}), (\ref{eq:t2}); $\rs$ drives $\zeta_2$ toward $\hat d$: the
aircraft yaws with the airmass at zero roll rate.

The improvement over disturbance-attenuating backstepping, the same control law
with $\hat d\equiv0$, follows from the transfers from $d=\dot\beta_v$ to
$\beta$, with $a=Y_\beta/V_T$:
\begin{subequations}\label{eq:tf}
\begin{eqnarray}
\text{attenuating:}\quad \beta&=&
\dfrac{-(s+c_1+c_2+a)}{(s+c_1)(s+c_2)+1}\;d \label{eq:tfwo}\\
\text{BDO:}\quad \beta&=&
\dfrac{-s\,(s+k+c_1+c_2+a)}{(s+k)\big[(s+c_1)(s+c_2)+1\big]}\;d \label{eq:tfw}
\end{eqnarray}
\end{subequations}
For the same $d$, the two sideslip signals stand in the ratio
\begin{equation}
\dfrac{\beta_{\rm BDO}(j\omega)}{\beta_{\rm att}(j\omega)}
=\dfrac{j\omega\,(j\omega+k+c_1+c_2+a)}{(j\omega+k)(j\omega+c_1+c_2+a)}
\ \longrightarrow\ \dfrac{j\omega}{j\omega+c_1+c_2+a}\quad (k\to\infty).
\label{eq:ratio}
\end{equation}
The DO places a zero at the origin: constant vortex rate is rejected
exactly. At the vortex rate $\omega_L$ the sideslip is smaller by about
$\omega_L/(c_1+c_2)$, an order of magnitude for $\omega_L\sim1\,{\rm s}^{-1}$
and a roll--yaw loop of several rad/s.

A higher-order DO, whose residual is driven by
$\dddot\beta_v$, gives more, at one more effective differentiation of the
vane signal per order. Closing the chain of derivatives instead of truncating
it is output regulation, Section~\ref{sec:or}.

\subsection{LQ with disturbance observer (LQ+DO)}\label{sec:lqdo}

We use the same $\hat d$ in the same way, as the operating point, with a
Riccati gain in place of the cancellation. For a constant $d$ the steady state of
(\ref{eq:AB}) with $\beta=\ps=0$ is
\begin{equation}
x_{\rm ss}(d)=\begin{pmatrix}0\\0\\1\end{pmatrix}\dfrac{d}{b_r},\qquad
u_{\rm ss}(d)=-\begin{pmatrix}L_{\delta_\ail}&L_{\delta_\rud}\\
N_{\delta_\ail}&N_{\delta_\rud}\end{pmatrix}^{-1}\begin{pmatrix}L_{\rs}\\ N_{\rs}\end{pmatrix}\dfrac{d}{b_r},
\qquad b_r=\dfrac{Y_{\rs}}{V_T}-1 ,
\label{eq:ss}
\end{equation}
and with $K=R^{-1}B^{\!\top}P$, $P$ the solution of
$A^{\!\top}P+PA-PBR^{-1}B^{\!\top}P+Q=0$,
\begin{equation}
\boxed{\;u=-K\big(x-x_{\rm ss}(\hat d)\big)+u_{\rm ss}(\hat d)\;}
\label{eq:lq}
\end{equation}
\begin{remark}\label{rem:comp}
In the measured states, with (\ref{eq:dobtf}),
(\ref{eq:ss}) substituted, (\ref{eq:lq}) is the compensator
\begin{subequations}\label{eq:comp}
\begin{eqnarray}
\begin{pmatrix}\delta_\ail\\ \delta_\rud\end{pmatrix}&=&\big[-K+G\,H(s)\big]
\begin{pmatrix}\beta\\ \ps\\ \rs\end{pmatrix} \label{eq:compC}\\
H(s)&=&\dfrac{k}{s+k}\Big(\dfrac{Y_\beta}{V_T}-s,\ \ \dfrac{Y_{\ps}}{V_T},\ \
\dfrac{Y_{\rs}}{V_T}-1\Big) \label{eq:compH}\\
G&=&\dfrac{V_T}{Y_{\rs}-V_T}\left[\begin{pmatrix}K_{13}\\K_{23}\end{pmatrix}
-\dfrac{1}{\Delta}\begin{pmatrix}N_{\delta_\rud}L_{\rs}-L_{\delta_\rud}N_{\rs}\\
L_{\delta_\ail}N_{\rs}-N_{\delta_\ail}L_{\rs}\end{pmatrix}\right] \label{eq:compG}\\
\Delta&=&L_{\delta_\ail}N_{\delta_\rud}-L_{\delta_\rud}N_{\delta_\ail} \label{eq:compD}
\end{eqnarray}
\end{subequations}
of dynamic order one: the static gain $-K$ and a rank-one term with the
single filter $k/(s+k)$. Since $H(s)x=\hat d$ by (\ref{eq:dobtf}),
(\ref{eq:compC}) is $u=-Kx+G\hat d$, and the BDO control law (\ref{eq:law}) has
the same form: with $c^{\!\top}=(Y_\beta/V_T,\,Y_{\ps}/V_T,\,Y_{\rs}/V_T-1)$
the row of (\ref{eq:z2}), so that $a_\zeta^{\!\top}=c^{\!\top}A$ and
$a_\eta^{\!\top}=e_2^{\!\top}A$, (\ref{eq:law}) is (\ref{eq:compC}) with
\begin{equation}
K=D^{-1}\begin{pmatrix}c^{\!\top}\big(A+(c_1+c_2)I\big)+(1+c_1c_2)\,e_1^{\!\top}\\[3pt]
e_2^{\!\top}\big(A+c_3I\big)\end{pmatrix},\qquad
G=D^{-1}\begin{pmatrix}c_1+c_2+Y_\beta/V_T\\ 0\end{pmatrix}
\label{eq:Kbdo}
\end{equation}
in place of the Riccati gain and (\ref{eq:compG}).
\end{remark}

With
$\tilde x=x-x_{\rm ss}(\hat d)$ and $e=\hat d-d$,
\begin{equation}
\dot{\tilde x}=(A-BK)\,\tilde x+\big(k\,x_{\rm ss}(1)-e_1\big)\,e,\qquad
\dot e=-k\,e-\ddot\beta_v ,
\label{eq:lqerr}
\end{equation}
the counterpart of (\ref{eq:target}): driven by the DO error alone,
with a coefficient that grows with $k$ as in (\ref{eq:t2}). The attenuating
version is $u=-Kx$. The full-information version replaces $\hat d$ by $d$ and
adds $\big[\begin{smallmatrix}L_{\delta_\ail}&L_{\delta_\rud}\\
N_{\delta_\ail}&N_{\delta_\rud}\end{smallmatrix}\big]^{-1}(0,\dot d/b_r)^{\!\top}$,
under which $\dot{\tilde x}=(A-BK)\tilde x$ and sideslip is rejected exactly.

BDO and LQ+DO use $\hat d$ identically and neither uses $\dot d$; the
difference between them is the feedback around the feedforward, and it is a
question of margins, taken up in Section~\ref{sec:margins}.

The LQ gains of Condition~II (Table~\ref{tab:cond}, Section~\ref{sec:forced}),
used in Sections~\ref{sec:margins} and \ref{sec:persist}, are
\begin{eqnarray}
&&Q={\rm diag}\big((0.5^\circ)^{-2},(3^\circ/{\rm s})^{-2},(5^\circ/{\rm s})^{-2}\big),\qquad
R={\rm diag}\big((10^\circ)^{-2},(10^\circ)^{-2}\big),\nonumber\\
&&K=\begin{pmatrix}4.11&-3.16&-1.79\\ 18.8&0.36&-6.94\end{pmatrix},
\label{eq:Kval}
\end{eqnarray}
with closed-loop poles $-18.3$, $-2.9\pm2.6j$. The backstepping gains used
there are $c_1=2$, $c_2=8$, $c_3=4$, for which (\ref{eq:Kbdo}) gives
\begin{equation}
K=\begin{pmatrix}7.35&-0.65&-3.87\\ 19.6&-0.23&-12.1\end{pmatrix},\qquad
G=\begin{pmatrix}3.46\\ 12.5\end{pmatrix},
\label{eq:Kbdoval}
\end{equation}
with closed-loop poles $-2.2$, $-7.8$, $-4$, the roots of
$(s+c_1)(s+c_2)+1$ and $-c_3$. The two state feedbacks are different: a
backstepping gain (\ref{eq:Kbdo}) has three free parameters against six in
an LQ gain, so an LQ gain is a backstepping gain only when it has the
structure (\ref{eq:Kbdo}), which (\ref{eq:Kval}) does not.

\subsection{LQR PI servo with feedforward}\label{sec:servo}

\begin{figure}[t]
\centering
\begin{tikzpicture}[>=Latex,node distance=6mm and 7mm,
 blk/.style={draw,minimum height=7mm,minimum width=9mm,inner sep=2pt},
 sm/.style={draw,circle,inner sep=0pt,minimum size=4.5mm},
 lab/.style={font=\scriptsize}]
\node (yc) {$y_{\rm c}$};
\node[sm,right=7mm of yc] (s1) {};
\node[blk,right=of s1] (int) {$\tfrac{1}{s}$};
\node[blk,right=of int] (KI) {$-K_I$};
\node[sm,right=9mm of KI] (s2) {};
\node[blk,right=12mm of s2] (B) {$B$};
\node[sm,right=of B] (s3) {};
\node[blk,right=of s3] (P) {$(sI-A)^{-1}$};
\node[blk,right=11mm of P] (C) {$C_r$};
\node[sm,right=of C] (s4) {};
\node[right=7mm of s4] (y) {$y$};
\node[blk] (Kc) at ($(KI)+(0,11mm)$) {$K_c$};
\node[blk] (Kd) at ($(s2)+(0,22mm)$) {$-K_d$};
\node[blk] (Bd) at ($(s3)+(0,22mm)$) {$B_d$};
\node[blk] (Dr) at ($(C)+(0,11mm)$) {$D_r$};
\node[blk] (Drd) at ($(s4)+(0,22mm)$) {$D_{rd}$};
\node (d) at ($(s3)+(0,33mm)$) {$\beta_v$};
\node[blk] (Kx) at ($(P)+(0,-11mm)$) {$-K_{x_p}$};
\draw[->] (yc)--(s1); \node[lab] at ($(s1.west)+(-1.5mm,2mm)$) {$-$};
\draw[->] (s1)--node[lab,above]{$e_I$}(int);
\draw[->] (int)--(KI); \draw[->] (KI)--(s2);
\coordinate (tc) at ($(yc)!0.45!(s1)$); \fill (tc) circle (0.6pt);
\draw[->] (tc)|-(Kc); \draw[->] (Kc)-|($(s2.north west)+(0.3mm,-0.3mm)$);
\draw[->] (s2)--(B); \coordinate (tu) at ($(s2.east)!0.5!(B.west)$); \fill (tu) circle (0.6pt);
\node[lab,above] at ($(tu)+(1.5mm,0)$) {$u$};
\draw[->] (tu)|-(Dr); \draw[->] (Dr)-|($(s4.north west)+(0.4mm,-0.4mm)$);
\draw[->] (B)--(s3); \draw[->] (s3)--(P);
\draw[->] (P)--(C); \coordinate (tx) at ($(P.east)!0.5!(C.west)$); \fill (tx) circle (0.6pt);
\node[lab,above] at (tx) {$x_p$};
\draw[->] (tx)|-(Kx); \draw[->] (Kx)-|(s2);
\draw[->] (C)--(s4); \draw[->] (s4)--(y);
\draw[->] (d)--(Bd); \draw[->] (Bd)--(s3);
\coordinate (td) at ($(d.south)+(0,-2mm)$);
\draw[->] (d.west)-|(Kd); \draw[->] (Kd)--(s2);
\draw[->] (d.east)-|(Drd); \draw[->] (Drd)--(s4);
\coordinate (ty) at ($(s4.east)!0.5!(y.west)$); \fill (ty) circle (0.6pt);
\draw[->] (ty)--++(0,-21mm)-|(s1);
\end{tikzpicture}
\caption{LQR PI servo with feedforward on the command and on the vortex
sideslip, (\ref{eq:servo}).}
\label{fig:servo}
\end{figure}
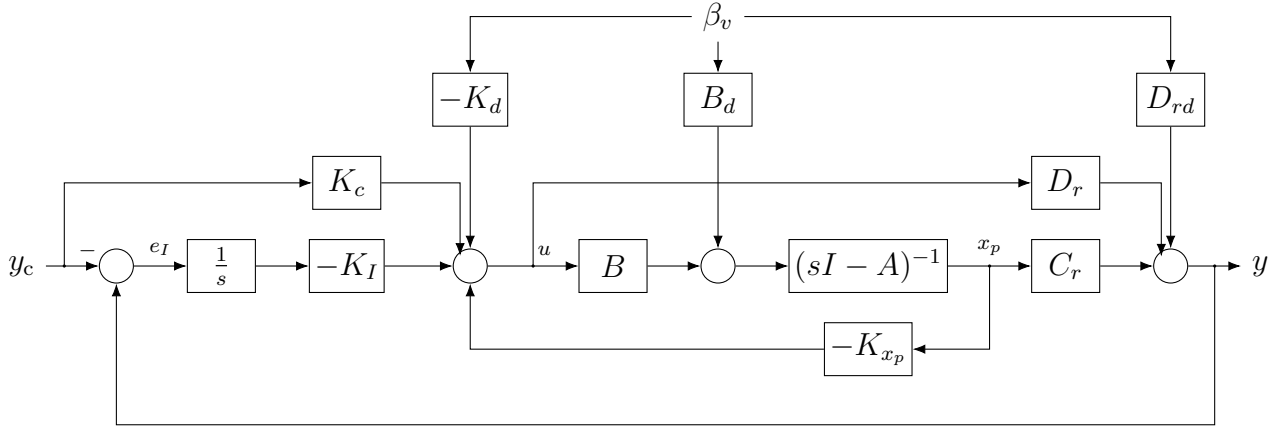

The servo-controller is designed using the system state with inertial
components $x_p=(\beta_b,\ps,\rs)$, in
which (\ref{eq:beta})--(\ref{eq:r}) read
\begin{equation}
\dot x_p=A\,x_p+B\,u+B_d\,\beta_v,\qquad
B_d=-A\,e_1=-\Big(\dfrac{Y_\beta}{V_T},\,L_\beta,\,N_\beta\Big)^{\!T},
\label{eq:inertial}
\end{equation}
where the vortex sideslip enters all three equations, through the side
force, the roll and yaw moments, by subtracting the $\beta$ column of $A$,
which in turn represents the effectiveness of the vortex sideslip $\beta_v$.
For command tracking, the regulated output is the stability-axis roll rate
and the lateral load factor,
\begin{equation}
y_r=\begin{pmatrix}\ps\\ N_y\end{pmatrix}=C_rx_p+D_ru+D_{rd}\,\beta_v,
\label{eq:yr}
\end{equation}
with $D_{rd}=-C_re_1$ since the airmass-related sideslip angle is
$\beta=\beta_b-\beta_v$. Adding the integrator dynamics
$\dot e_I=y_r-y_{\rm c}$, (\ref{eq:inertial}) extends to
\begin{equation}
\begin{pmatrix}\dot e_I\\ \dot x_p\end{pmatrix}=
\begin{pmatrix}0&C_r\\0&A\end{pmatrix}\begin{pmatrix}e_I\\ x_p\end{pmatrix}
+\begin{pmatrix}D_r\\ B\end{pmatrix}u
+\begin{pmatrix}D_{rd}\\ B_d\end{pmatrix}\beta_v
+\begin{pmatrix}-I\\ 0\end{pmatrix}y_{\rm c},
\label{eq:ext}
\end{equation}
This system remains controllable when the Rosenbrock matrix has no transmission zero at the
origin,
\begin{equation}
\det M\neq0,\qquad M=\begin{pmatrix}A&B\\ C_r&D_r\end{pmatrix}.
\label{eq:rosen}
\end{equation}
The LQR PI control law is $u=-K_Ie_I-K_{x_p}x_p$, with $(K_I,K_{x_p})$, computed with
the side forces of the surfaces retained in $B$ and $D_r$, using the Riccati gain of
(\ref{eq:ext}). The command feedforward follows from the equilibrium
$(x_p^{\rm eq},u^{\rm eq})=M^{-1}(0,I)^{\!\top}y_{\rm c}$ and
$u^{\rm eq}=-K_{x_p}x_p^{\rm eq}+K_cy_{\rm c}$,
\begin{equation}
K_c=\begin{pmatrix}K_{x_p}&I\end{pmatrix}M^{-1}\begin{pmatrix}0\\ I\end{pmatrix},
\label{eq:Kc}
\end{equation}
and the disturbance feedforward $u=-K_d\beta_v$ minimizes the contribution
of the airmass vorticity term $\big|(D_r;B)\,u+(D_{rd};B_d)\,\beta_v\big|$
on the system dynamics,
\begin{equation}
K_d=\big(D_r^{\!\top}D_r+B^{\!\top}B\big)^{-1}
\begin{pmatrix}D_r\\ B\end{pmatrix}^{\!\top}\begin{pmatrix}D_{rd}\\ B_d\end{pmatrix}.
\label{eq:Kd}
\end{equation}
The servo control law is (Figure~\ref{fig:servo})
\begin{equation}
\boxed{\;u=-K_Ie_I-K_{x_p}x_p+K_cy_{\rm c}-K_d\beta_v\;}
\label{eq:servo}
\end{equation}
The feedforward $K_d\beta_v$ cancels the roll and yaw moments of $\beta_v$
(exactly when the side forces of the surfaces are dropped), not its effect
on $\beta=\beta_b-\beta_v$, on which $\beta_v$ acts directly, without an
integrator. The design regulates $\ps$ with small deviations in $N_y$ and $\beta$:
a coordinated roll maneuver.

With $\beta_v$ unknown, integrate the DO estimate $\hat d$ according to
$\dot{\hat\beta}_v=\hat d$, with $\hat\beta_v(0)=\beta_v(0)$; in the test of
Section~\ref{sec:trackff} both are zero:
\begin{equation}
\hat\beta_v=\frac{1}{s}\,\hat d,
\label{eq:bvhat}
\end{equation}
which is $\frac{k}{s+k}\beta_v$ by (\ref{eq:dobd}) for zero initial
estimation error, and (\ref{eq:servo}) is applied as
\begin{equation}
\boxed{\;u=-K_Ie_I-K_{x_p}\big(x+e_1\hat\beta_v\big)+K_cy_{\rm c}-K_d\hat\beta_v\;}
\label{eq:servodo}
\end{equation}
with $e_I$ the integral of the measured $y_r-y_{\rm c}$. The DO input
$\frac{g}{V_T}N_y-\rs$ is needed here: with $\zeta_2$ in its place,
$\hat\beta_v$ is offset by the integral of $\frac{k}{s+k}\zeta_u$, which
drifts whenever the surfaces hold a deflection (Figure~\ref{fig:dhat}).

\subsection{Stability robustness margins}\label{sec:margins}

We compute loop-at-a-time margins of the linear closed loop, the plant
(\ref{eq:AB}) with the DO (\ref{eq:dobtf}) and either control law, exosystem
removed. Gain ranges are exact, from the closed-loop eigenvalues with a gain inserted in one
channel; the phase margins are the disk bound $2\arcsin(s_{\min}/2)$,
$s_{\min}=\min_\omega|1+\ell(j\omega)|$ over the loop $\ell$ broken at that
channel with the others closed.

\begin{figure}[t]
\centering
\begin{tikzpicture}[>=Latex,node distance=6mm and 8mm,
 blk/.style={draw,minimum height=7mm,minimum width=10mm,inner sep=3pt},
 sm/.style={draw,circle,inner sep=0pt,minimum size=4.5mm},
 lab/.style={font=\scriptsize}]
\node (yc) {$y_{\rm c}$};
\node[sm,right=6mm of yc] (s1) {};
\node[blk,right=of s1] (pi) {PI servo};
\node[sm,right=of pi] (s2) {};
\node[right=12mm of s2] (brk) {$\boldsymbol\times$};
\node[blk,right=8mm of brk] (act) {$G_{\rm act}(s)$};
\node[blk,right=10mm of act] (pl) {plant};
\node[right=8mm of pl] (yr) {$y_r$};
\node[blk] (do) at ($(s2)+(0,14mm)$) {DO};
\node (bv) at ($(pl)+(0,14mm)$) {$\beta_v$};
\draw[->] (yc)--(s1); \node[lab] at ($(s1.west)+(-1.5mm,2mm)$) {$-$};
\draw[->] (s1)--(pi); \draw[->] (pi)--(s2);
\draw[->] (s2)--node[lab,above,pos=0.2]{$u$}(brk); \draw[->] (brk)--(act);
\draw[->] (act)--(pl); \draw[->] (pl)--(yr);
\draw[->] (bv)--(pl);
\draw[->] (do)--(s2);
\coordinate (ty) at ($(pl.east)!0.5!(yr.west)$); \fill (ty) circle (0.6pt);
\draw[->] (ty)--++(0,-9mm)-|(s1);
\draw[->] (pl.north west)++(1mm,0)|-++(0,22mm)coordinate(tx)-|(do.north);
\node[lab,above] at ($(tx)!0.5!(do.north|-tx)$) {$x$, $N_y$};
\draw[->] ($(do.north|-tx)$)-|(pi.north);
\end{tikzpicture}
\caption{Loop of the servo (\ref{eq:servodo}) with the DO, broken at the
actuator command $u$ ($\times$) for the input margins.}
\label{fig:break}
\end{figure}
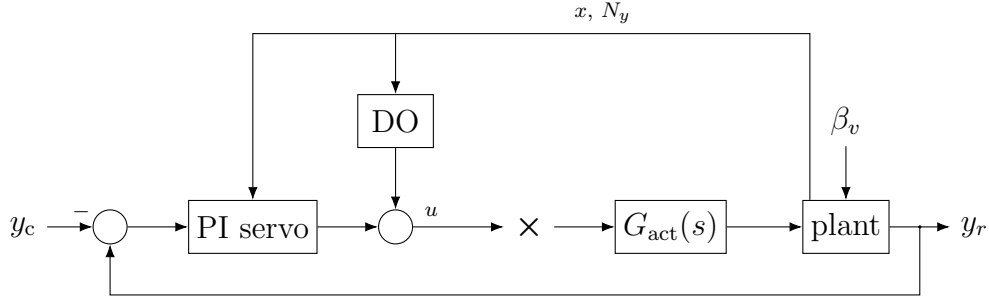

The DO is a feedback, through (\ref{eq:dobtf}), and it is included in the
loop gains; at the input channels its contribution is zero. Along motion
driven by the inputs alone, with the side forces dropped,
$\zeta_2=\dot\zeta_1$ identically, so
$\hat d=\frac{k}{s+k}(\zeta_2-s\zeta_1)\equiv0$, that is, $H(s)P(s)\equiv0$
with $P(s)=(sI-A)^{-1}B$, and the loop gain broken at either input is the
same with and without the DO. The input margins of LQ+DO are therefore
those of the LQR, and the return-difference inequality gives $s_{\min}\ge1$
at each input: gain $(1/2,\infty)$ and phase $60^\circ$, guaranteed at every
design point. The same holds for BDO. The margins are in
Table~\ref{tab:margins}. For the LQR PI servo (\ref{eq:servodo}), broken
at the actuator command $u$ (Figure~\ref{fig:break}), the DO contributes
nothing to the loop gain: with the measured $\frac{g}{V_T}N_y-\rs$,
(\ref{eq:dobd}) holds for any $u$, so $\hat\beta_v$ does not respond to a
signal injected at the break point, and the input margins, integrators and
actuators included, are the same with and without the DO
(Section~\ref{sec:trackff}): SISO, each loop with the other closed; MIMO, both
loops at once, from $\min_\omega\sigma_{\min}(I+L)=0.93$.

\begin{table}[t]
\caption{Loop-at-a-time margins, Condition~II of Table~\ref{tab:cond}, exosystem removed: gain range, then disk phase margin.}
\label{tab:margins}
\centering
\begin{tabular}{lccc}
\toprule
loop broken at & LQ & LQ+DO & BDO\\
\midrule
input $\delta_\ail$ & $[0,\infty)$, $\ge60^\circ$ & $[0,\infty)$, $\ge60^\circ$ & $[0,\infty)$, $\ge60^\circ$\\
input $\delta_\rud$ & $[0,\infty)$, $\ge60^\circ$ & $[0,\infty)$, $\ge60^\circ$ & $[0,\infty)$, $\ge60^\circ$\\
output $\beta$      & $[0,\infty)$, $\ge48^\circ$ & $[0,\infty)$, $\ge50^\circ$ & $[0,\infty)$, $\ge50^\circ$\\
output $\ps$        & $[0,\infty)$, $\ge60^\circ$ & $[0,\infty)$, $\ge60^\circ$ & $[0,\infty)$, $\ge60^\circ$\\
output $\rs$        & $[0,\infty)$, $\ge60^\circ$ & $[0,\ 32\,{\rm dB}]$, $\ge56^\circ$ & $[0,\ 32\,{\rm dB}]$, $\ge56^\circ$\\
\bottomrule
\end{tabular}
\end{table}

The ranges in Table~\ref{tab:margins} extend to zero
because either surface alone stabilizes this plant. The DO enters the
output loops and imposes a finite upper gain limit, $32$~dB, on the yaw-rate
measurement, through which $\zeta_2$ feeds $\hat d$. On this plant at these
gains BDO's margins are as large as LQ's; the difference is that the LQR
margins are guaranteed by the return-difference inequality at every one of
the flight points a production design covers, while backstepping's are
whatever the assigned poles produce and have to be checked point by point.
Margins of LQ, LQ+DO and BDO with the side forces retained have not been computed.

\section{Trade study: attenuation by LQ+DO vs.\ LQ alone}\label{sec:trade}

Sections~\ref{sec:tensec}, \ref{sec:sweep}, \ref{sec:persist} and
\ref{sec:doatt} present a control regulation design with respect to the
system equilibrium. Specifically, a trade study is conducted using LQ-based
proportional feedback controllers with and without a disturbance observer.
In Section~\ref{sec:track}, an LQR PI servo-controller is designed to track
the aircraft roll-rate and lateral acceleration commands, both with and
without the DO option.

\subsection{Ten-second transient at resonance}\label{sec:tensec}

The roll/yaw model represents the roll and dutch-roll modes and is intended
for short-duration use, on the order of several seconds, so the closed-loop
test is limited to a ten-second transient. The
gains, inside the $4$--$6$~rad/s window of Section~\ref{sec:sweep}, are $R$
as in (\ref{eq:Kval}) and
$Q={\rm diag}\big((1^\circ)^{-2},(20^\circ/{\rm s})^{-2},(20^\circ/{\rm s})^{-2}\big)$,
which give poles $-3.4\pm3.6j$, $-0.56$ and the gain (\ref{eq:Khigh});
$k=20$, and $\alpha=30^\circ$ in $p_b$.
At $\omega_n$ (Figure~\ref{fig:freq}) LQ attenuation gives $-4.6$~dB and LQ+DO
$-12.7$~dB, eight decibels apart, and both control laws move the peak of the transfer
to the loop bandwidth, $+2$~dB at $6$~s$^{-1}$ for LQ and at $10$~s$^{-1}$
for LQ+DO. The eight-decibel improvement is consistent with the mechanism exhibited
by (\ref{eq:ratio}): the vortex rate is below the loop bandwidth.

\begin{figure}[t]
\centering
\includegraphics[width=0.86\textwidth,trim=0 0 0 261,clip]{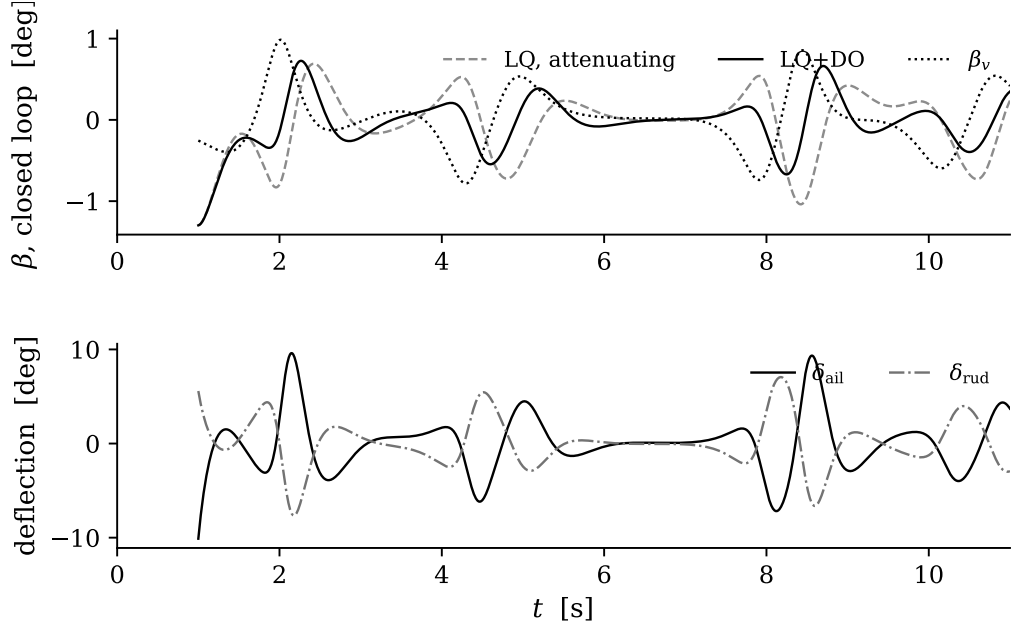}
\caption{F-16 at $\alpha=30^\circ$, $50$~psf, cell vortex: closed loop after
a $2^\circ$ aileron pulse, LQ and LQ+DO, with deflections.}
\label{fig:cl}
\end{figure}

The closed-loop runs (Figure~\ref{fig:cl}) last ten seconds and start from the state at the end
of a $2^\circ$ aileron pulse of one second, $\beta=-1.1^\circ$,
$\ps=-2.4^\circ$/s, $\rs=1.3^\circ$/s, the transient that recurs in flight,
with the Lorenz state on the attractor at twenty different phases. Without the vortex the two control laws coincide, since $\hat d\equiv0$ then. With
the vortex, after the two-second transient, over the twenty runs, the result is
Table~\ref{tab:tensec}.
\begin{table}[t]
\caption{Ten-second transient, Condition~I of Table~\ref{tab:cond}, twenty attractor phases, after the two-second transient.}
\label{tab:tensec}
\centering
\begin{tabular}{lccc}
\toprule
& vortex $\beta_v$ & $\beta$, LQ attenuating & $\beta$, LQ+DO\\
\midrule
rms, mean of runs & $0.36^\circ$ & $0.35^\circ$ & $0.25^\circ$\\
peak, mean of runs & --- & $0.94^\circ$ & $0.71^\circ$\\
\bottomrule
\end{tabular}
\end{table}
The ratio is $0.70$, the same at $\zeta=0.08$, and the peak deflections are
$12^\circ$ aileron and $10^\circ$ rudder. The ratio is above the $-8$~dB of Figure~\ref{fig:freq}
because the vortex sideslip is not a single frequency at $\omega_n$: the circulation of the origin
through the cell, at a period of a few seconds, spreads its spectrum to
frequencies where the two control laws differ less.

\begin{figure}[t]
\centering
\includegraphics[width=0.8\textwidth]{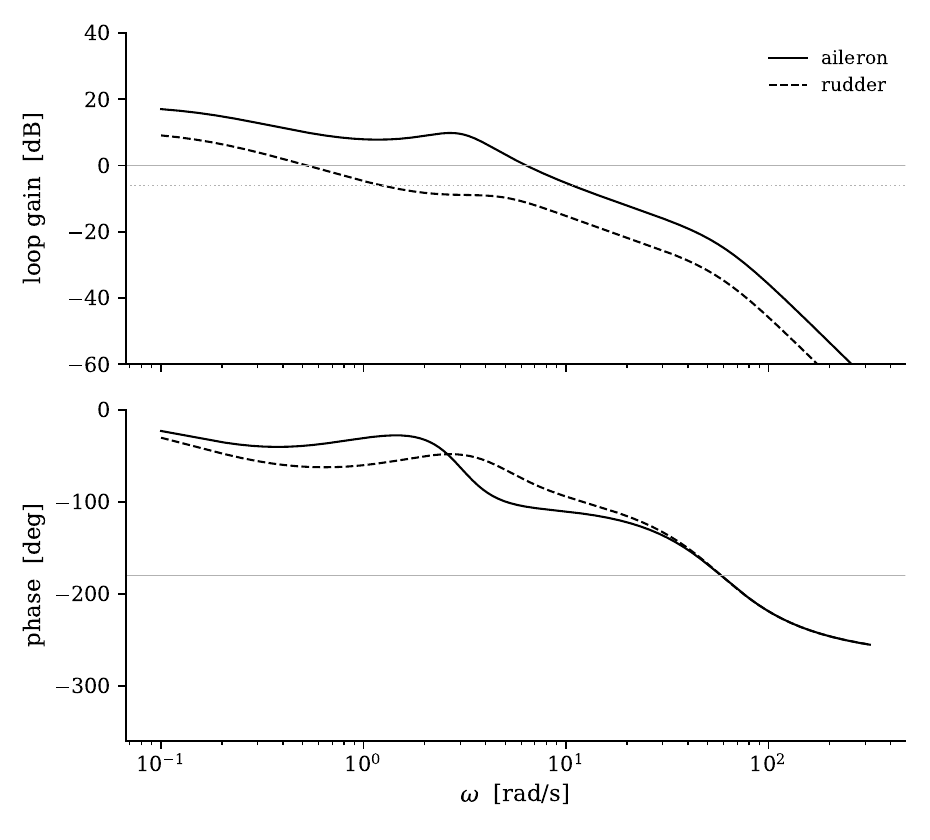}
\caption{Loop gains broken at the aileron and rudder commands, other loop
closed, with actuators, $Q_\beta=(1^\circ)^{-2}$, $k=20$; identical for LQ
and LQ+DO.}
\label{fig:bode}
\end{figure}
\begin{figure}[t]
\centering
\includegraphics[width=0.8\textwidth]{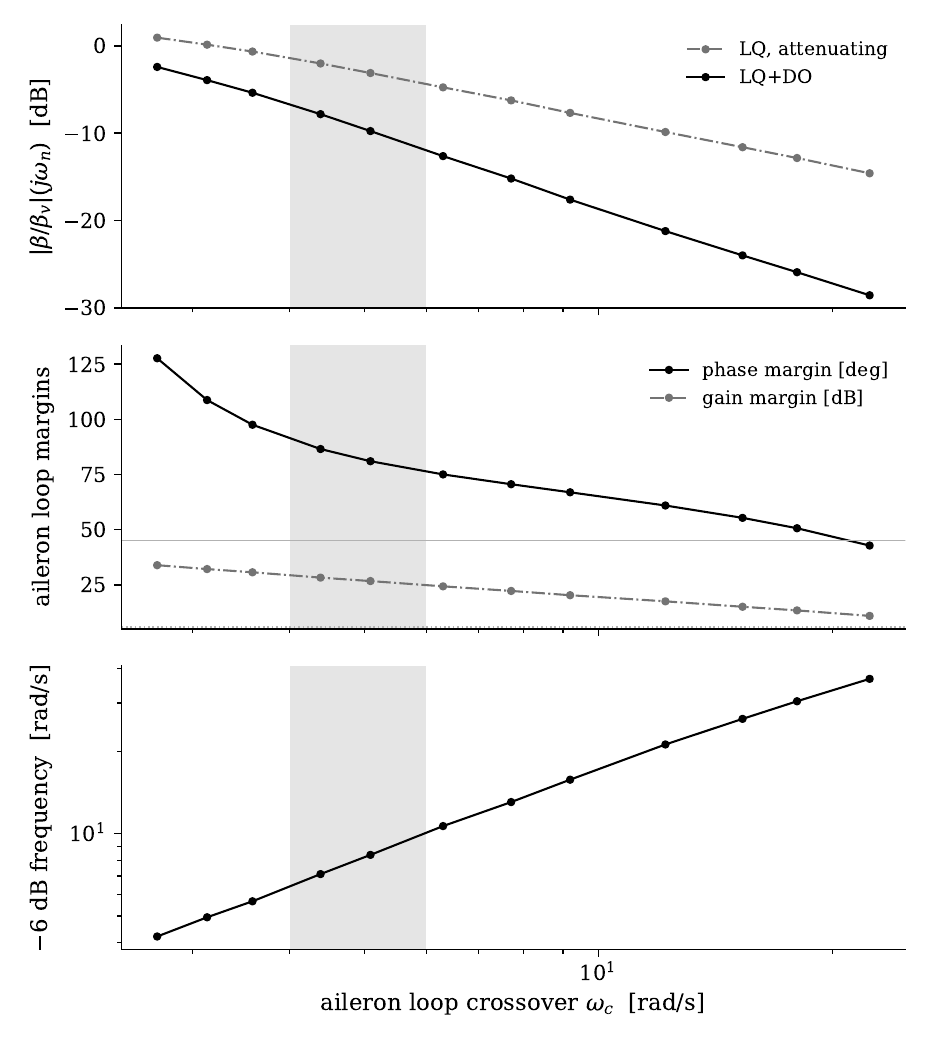}
\caption{Sweep of $Q_\beta$ with actuators against aileron-loop crossover:
vortex-to-sideslip gain at the dutch roll, margins, and $-6$~dB frequency;
shaded, $4$--$6$~rad/s.}
\label{fig:sweep}
\end{figure}

\subsection{Actuators, bandwidth sweeps and loop gains}\label{sec:sweep}

The constraints are for the analysis, not the design, as in flight control
practice: second-order
actuators $\omega_a^2/(s^2+2\zeta_a\omega_as+\omega_a^2)$ on both surfaces,
$\zeta_a=0.7$, $\omega_a=60$~rad/s; a closed-loop bandwidth of $4$--$6$~rad/s
for a tactical aircraft, anything above $1$~Hz being high gain; and
attenuation of the first structural mode by $6$~dB. The design has two
parameters, the LQ weight $Q_\beta$, which sets the loop bandwidth, and the
DO bandwidth $k$; each is swept with the other fixed, at
$\alpha=30^\circ$, $50$~psf, actuators in. At each point we record the
vortex-to-sideslip gain at the dutch roll $\omega_n=2.09$, the margins of the loops broken at
the aileron and at the rudder command with the other loop closed, and the
frequency above which each loop gain stays below $-6$~dB, to be compared
with whatever the first structural mode is.

Figure~\ref{fig:bode} shows the two loop gains at $Q_\beta=(1^\circ)^{-2}$,
$k=20$: aileron crossover $6.3$~rad/s, phase margin $75^\circ$, gain margin
$24$~dB, below $-6$~dB above $10.7$~rad/s; the rudder loop is a low-gain
loop, crossover $0.5$~rad/s, margins $118^\circ$ and $34$~dB. Over the
sweep of Figure~\ref{fig:sweep}, aileron crossover $3.6$ to $32$~rad/s, the
rudder crossover stays at $0.49$--$0.55$~rad/s, phase margin
$117$--$120^\circ$. The plots are
the same for LQ attenuation and LQ+DO, and the same for every $k$: the
DO is absent from the loops broken at the commands, actuators
included, as in Section~\ref{sec:margins}.

The $2\times2$ loop gain at the control input, negative feedback with the
minus sign outside, is
\begin{equation}
L(s)=\big[K-G\,H(s)\big]\,P_a(s)\,P(s),\qquad
P(s)=(sI-A)^{-1}B,\qquad P_a(s)=\dfrac{\omega_a^2}{s^2+2\zeta_a\omega_as+\omega_a^2},
\label{eq:loopgain}
\end{equation}
$P(s)$ the $3\times2$ plant from the surfaces to $(\beta,\ps,\rs)$. The
absence of the DO is an identity:
$H(s)P(s)\equiv0$, because $(Y_\beta/V_T-s)\beta+(Y_{\ps}/V_T)\ps
+(Y_{\rs}/V_T-1)\rs$ is the sideslip equation (\ref{eq:beta}) with the vortex sideslip
removed, which the input-driven motion satisfies exactly (side forces of the
surfaces dropped). So $L(s)=KP_a(s)P(s)$ for every $k$, and the input
margins are the LQR's.

Figure~\ref{fig:sweep} is the bandwidth sweep at $k=20$. Inside the
$4$--$6$~rad/s window LQ attenuation gives $-2$ to $-4$~dB at the dutch
roll, leaving most of the resonant response, and LQ+DO gives $-8$ to
$-12$~dB; the contribution of the DO, $6$ to $8$~dB, is obtained where
loop gain is not permitted. The margins are
$81$--$87^\circ$ and $27$--$28$~dB there, and the $-6$~dB frequency is
$7$--$8$~rad/s. Raising the crossover gives $3$~dB per doubling for
either control law; the phase margin passes $45^\circ$ at a crossover of
$22$~rad/s ($Q_\beta=(0.1^\circ)^{-2}$), where the $-6$~dB frequency has
risen to $37$~rad/s, into the range of structural modes. The method is
therefore valid up to a crossover of about $18$~rad/s on margins alone and to whatever
crossover the structural mode allows, and the $4$--$6$~rad/s window is a
factor of three below both.

In the DO sweep at $Q_\beta=(0.5^\circ)^{-2}$ the gain at the dutch roll
is $-9$~dB at $k=2$, $-12$ at $5$, $-15$ at $10$, $-17.6$ at $20$, $-19$ at
$40$, $-20$ at $80$ and $-20.7$ at $320$. It saturates at two to four times
the loop crossover, at the corresponding finite-bandwidth limit, and the margins at the commands do not move with $k$. What $k$ does
change is the transfer from sensor noise to the surfaces, through the direct
term $-kG$ of (\ref{eq:compH}) on $\beta$; that, not the loop margins, is
what bounds $k$ in practice.

\subsection{Roll-rate command tracking}\label{sec:track}

The test is Condition~I with the cell vortex and actuators limited to rates
of $80^\circ$/s on the aileron and $120^\circ$/s on the rudder and to
deflections of $21.5^\circ$ and $30^\circ$. The command is a stability-axis
roll-rate doublet, $p_{\rm c}=10^\circ$/s from $1$ to $5$~s, $-10^\circ$/s
from $8$ to $12$~s and zero otherwise, with zero sideslip command: a
coordinated roll about the velocity vector.

\subsubsection{Sideslip attenuation}\label{sec:trackbeta}

The baseline is the LQR PI servo of \cite{LW}: the integrals of $\beta$ and $\ps-p_{\rm c}$ are
appended to $x$, and the Riccati gain of the five states is computed with
\begin{equation}
Q={\rm diag}(200,\,20,\,0,\,0,\,100),\qquad R={\rm diag}(4,\,1),
\label{eq:Qservo}
\end{equation}
in radians. The weights are tuned as in \cite{LW}: the integral weights set
the speed of tracking, $\ps$ is not weighted since it is the regulated
output and must move, the weight on $\rs$ sets the dutch-roll damping, and
the ratio of the entries of $R$ equalizes the two crossovers. The result is
aileron and rudder crossovers of $6.1$ and $6.0$~rad/s, phase margins of
$100^\circ$, and dutch-roll damping $0.72$.

The DO enters as in (\ref{eq:lq}): the servo control law is shifted to the operating
point $x_{\rm ss}(\hat d)$ of (\ref{eq:ss}), whose only nonzero entry is the
yaw rate $\rs=\hat d/b_r$. To hold $\beta$ at zero, the aircraft yaws with
the airmass, and every peak of the vortex rate is a demand on the rudder.
With the design of Section~\ref{sec:lqdo} as derived, we varied the DO
bandwidth $k$ from $5$ to $20$, the integral weights from $50$ to $800$ and
the weight on $\rs$ from $3$ to $100$: in no case did the rudder rate stay
within $120^\circ$/s. We therefore command the fraction $\sigma$ of the
yaw rate that follows the airmass, $x_{\rm ss}(\sigma\hat d)$, which adds
$\sigma G\hat d$ to the servo control law, with $G$ from (\ref{eq:compG}) with the
state gain of the servo in place of $K$. The factor $\sigma\in(0,1]$ is not
part of the derivation of Section~\ref{sec:lqdo}, which is $\sigma=1$, and
corresponds to no physical coefficient of the model; it is chosen by trial,
against the rudder rate limit. We use $\sigma=0.4$ and
$k=18\approx3\omega_c$, by (\ref{eq:kguide}).

With this design the DO performs its task, the attenuation of sideslip,
within the actuator limits (Figure~\ref{fig:track}). Over ten attractor
phases it reduces the sideslip rms from $0.34^\circ$ to $0.24^\circ$ and
the largest sideslip from $1.38^\circ$ to $0.96^\circ$; the rudder stays
within $18^\circ$ and is at its rate limit $0.2\%$ of the time, the aileron
never. Without the vortex the two control laws give the same response, since
$H(s)P(s)\equiv0$: $\ps$ reaches $90\%$ of the command in $1.3$~s and
$|\beta|<0.16^\circ$. The DO does not improve roll-rate tracking, which is
not its task: the vortex reaches $\ps$ through $L_\beta\beta$ in
(\ref{eq:p}), which is left to the feedback, and the roll-rate deviation due
to the vortex is $0.52^\circ$/s rms without the DO and $0.67^\circ$/s with
it.

With $\sigma=1$ the sideslip rms falls by $48\%$, but the rudder reaches its
$30^\circ$ stop, is at its rate limit $4.5\%$ of the time, and the roll-rate
deviation nearly triples.

\subsubsection{Servo with feedforward}\label{sec:trackff}

\begin{figure}[p]
\centering
\includegraphics[width=0.9\textwidth]{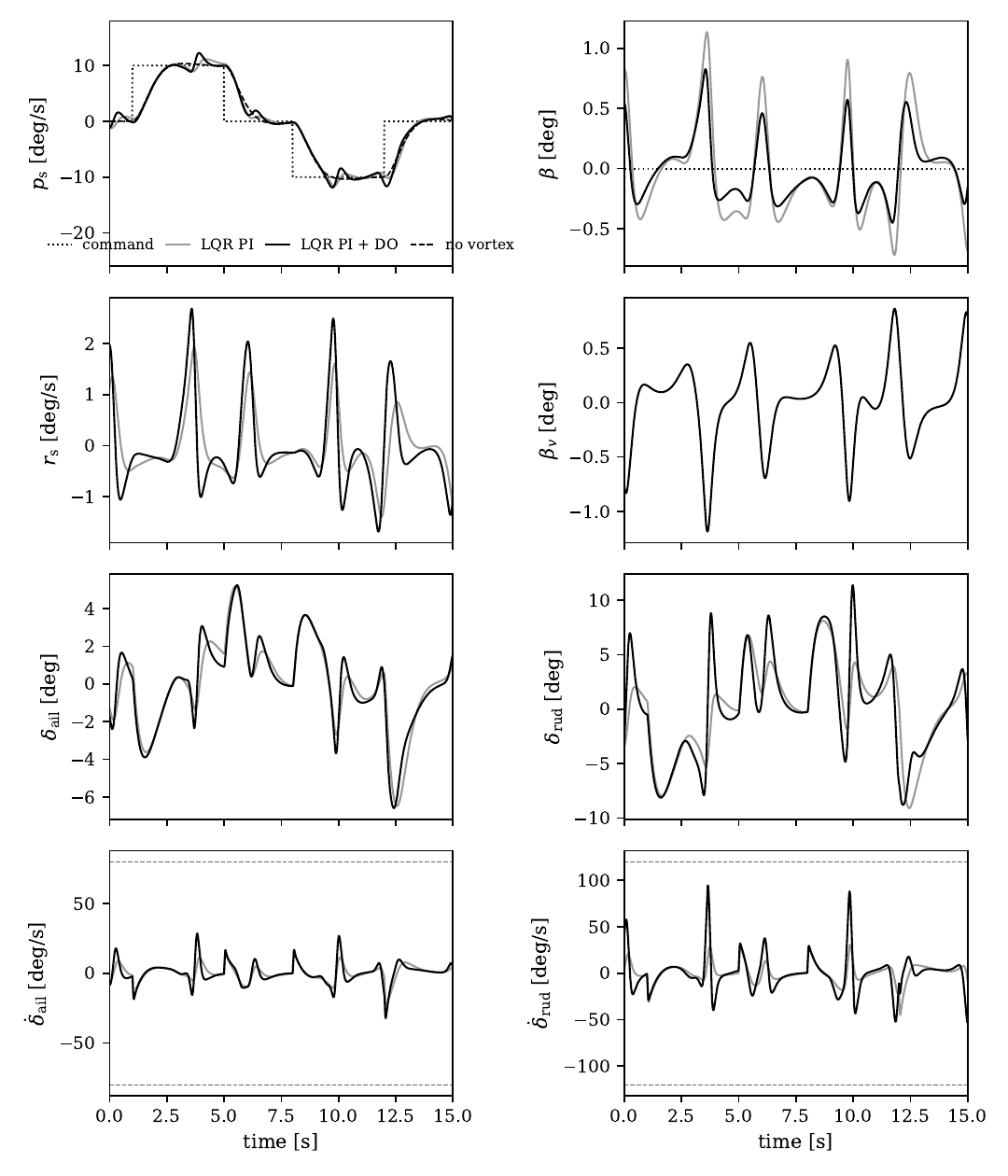}
\caption{Roll-rate doublet, Condition~I, cell vortex, rate-limited
actuators: LQR PI and LQR PI with DO, $\sigma=0.4$; dotted, commands and rate limits;
dashed, no vortex.}
\label{fig:track}
\end{figure}

The weights for $(e_I,x_p)$ are $Q={\rm diag}(1,1,0,0,1.5)$ and
$R={\rm diag}(1,0.5)$, $K_c$ is scaled by $0.2$ against overshoot, and
$k=18$. The test differs from that of Figure~\ref{fig:track} in the plant,
which retains the control side forces in (\ref{eq:beta}), in the vortex
and in the command: $F_0=F_0^{\rm crit}=0.03$, and at $t=0$
$(X,Y,Z)=(0,-0.001,0.001)$ with the airmass origin on the cell boundary,
$(y,z)=(0,-0.1)$~ft; $p_{\rm c}$ rises to $10^\circ$/s over $1$--$3$~s and
$20$--$22$~s, returns to zero over $10$--$12$~s and $30$--$32$~s, and is
passed through $1/(0.4s+1)$.

\begin{figure}[t]
\centering
\includegraphics[width=0.86\textwidth]{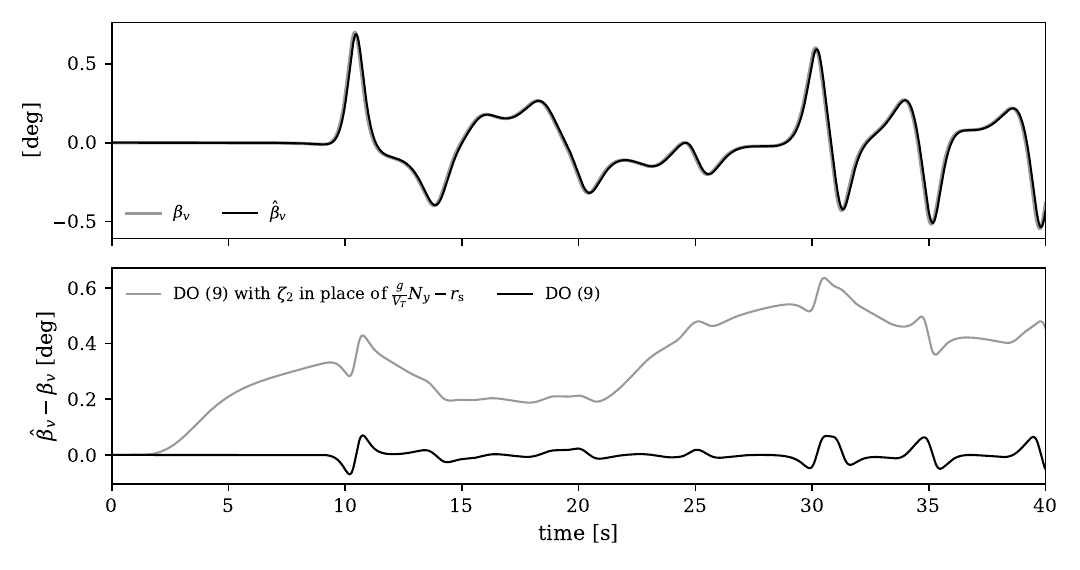}
\caption{Vortex sideslip $\beta_v$ and estimate $\hat\beta_v$, $k=18$, run
of Figure~\ref{fig:servoff}; lower panel, estimate error; grey, $\zeta_2$ in
place of $\frac{g}{V_T}N_y-\rs$.}
\label{fig:dhat}
\end{figure}

\begin{figure}[t]
\centering
\includegraphics[width=0.9\textwidth]{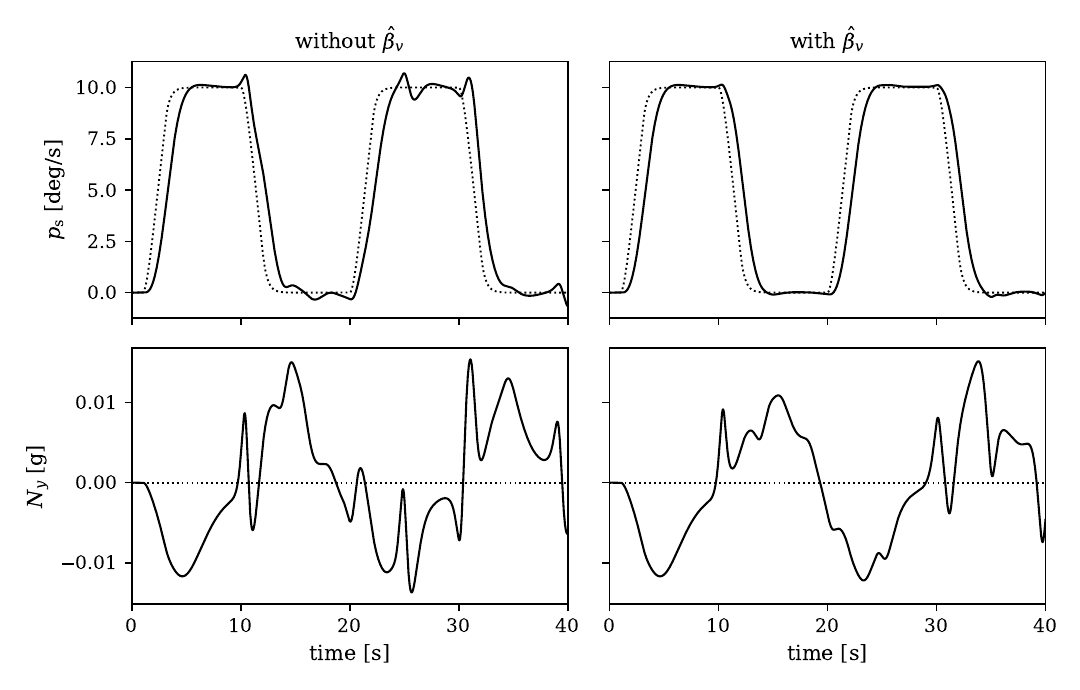}
\caption{Roll-rate doublets, servo (\ref{eq:servodo}): roll rate and lateral load factor without (left) and
with $\hat\beta_v$ (right); dotted, filtered command.}
\label{fig:servoff}
\end{figure}

The DO estimate (\ref{eq:bvhat}) follows the vortex sideslip with the lag
$k/(s+k)$, $\beta_v-\hat\beta_v=\hat d/k$; with $\zeta_2$ in place of
$\frac{g}{V_T}N_y-\rs$ in (\ref{eq:dobtf}), the estimate drifts during the rolls, by up to
$0.6^\circ$ in this test (Figure~\ref{fig:dhat}). Without
$\hat\beta_v$, the vortex sideslip reaches the roll rate through its roll
and yaw moments, and the feedback acts on them only through the motion
they produce: $\ps$ departs from the command each time the
vortex sideslip changes, during the rolls and between them
(Figure~\ref{fig:servoff}, left). With $\hat\beta_v$, the feedforward
$K_d\hat\beta_v$ cancels these moments as they arise, and $\ps$ follows the
command as it does without the vortex (Figure~\ref{fig:servoff}, right);
the peak lateral load factor is $0.015$~g in both runs. The feedforward is formed from
the measurements through the DO and uses no model of the airmass, and the
surfaces move less than $4^\circ$ (Table~\ref{tab:servo}).

\begin{figure}[t]
\centering
\includegraphics[width=0.8\textwidth]{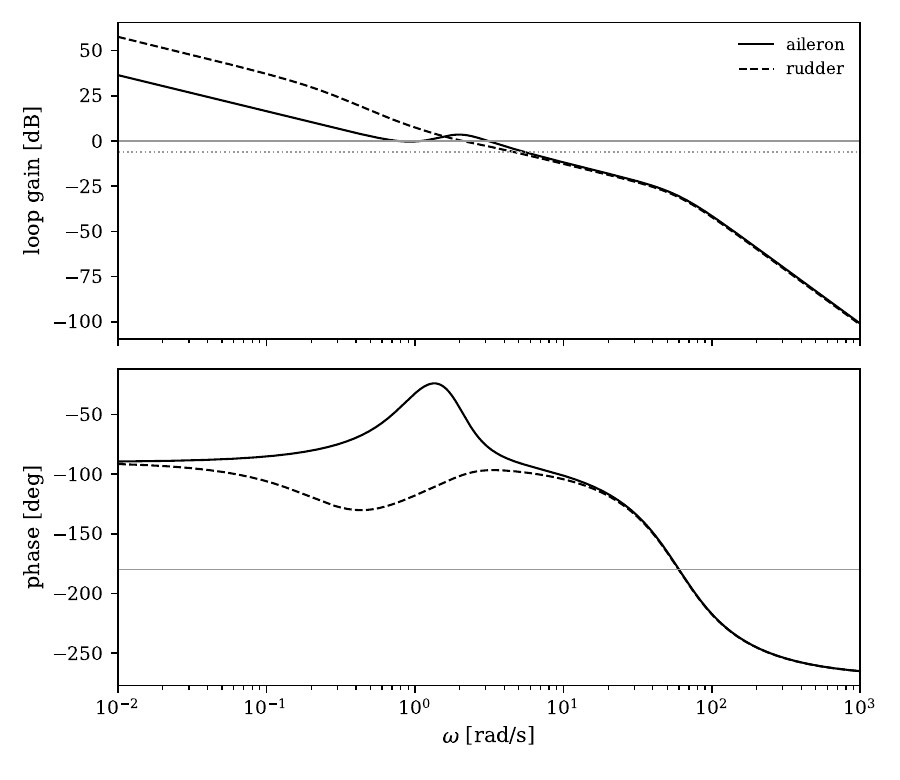}
\caption{Loop gains of the servo (\ref{eq:servodo}) at the aileron
and rudder commands, other loop closed, with actuators; the same with and
without the DO.}
\label{fig:bodeservo}
\end{figure}

Figure~\ref{fig:bodeservo} shows the loop gains of the servo at the
actuator command, each with the other loop closed, the same with and
without the DO. Below crossover the integrators give $-20$~dB/decade, and
the loop gains grow without bound as $\omega\to0$, which gives zero
steady-state error in $\ps$ and $N_y$; the aileron loop gain stays between
$-0.4$ and $3.5$~dB from $0.7$ to $3.2$~rad/s. Above crossover the loop
gains roll off at $-20$~dB/decade, and at $-60$~dB/decade above the
$60$~rad/s actuators; they are below $-6$~dB above $5.4$~rad/s (aileron)
and $4.5$~rad/s (rudder).

\begin{table}[t]
\centering
\caption{Servo (\ref{eq:servodo}) in the test of
Figure~\ref{fig:servoff}: peaks, and input margins at the actuator command,
the same with and without the DO.}
\label{tab:servo}
\renewcommand{\arraystretch}{1.15}
\setlength{\tabcolsep}{4pt}
\begin{tabular}{lccc}
\hline\hline
vortex & with $\hat\beta_v$ & without $\hat\beta_v$ & true $\beta_v$\\
\hline
roll-rate deviation due to the vortex & $0.15^\circ$/s & $1.11^\circ$/s & $0.13^\circ$/s\\
sideslip $\beta$ & $0.94^\circ$ & $1.12^\circ$ & $0.92^\circ$\\
$|N_y|$ & $0.015$~g & $0.015$~g & $0.016$~g\\
rudder, deflection and rate & $3.7^\circ$, $2.9^\circ$/s & $3.7^\circ$, $4.5^\circ$/s & $3.7^\circ$, $2.3^\circ$/s\\
aileron, deflection and rate & $2.3^\circ$, $2.6^\circ$/s & $2.2^\circ$, $3.2^\circ$/s & $2.3^\circ$, $2.7^\circ$/s\\
\hline
no vortex & \multicolumn{3}{c}{}\\
\hline
$|\beta|$, $|N_y|$ & \multicolumn{3}{c}{$0.92^\circ$, $0.012$~g}\\
\hline
input margins & \multicolumn{3}{c}{}\\
\hline
aileron, SISO: crossover, gain, phase & \multicolumn{3}{c}{$3.1$~rad/s, $30.6$~dB, $103^\circ$}\\
rudder, SISO: crossover, gain, phase & \multicolumn{3}{c}{$2.1$~rad/s, $31.1$~dB, $79^\circ$}\\
both loops, MIMO: gain, phase & \multicolumn{3}{c}{$-5.7/{+}23.1$~dB, $\pm55^\circ$}\\
\hline\hline
\end{tabular}
\end{table}

\begin{figure}[t]
\centering
\includegraphics[width=0.86\textwidth]{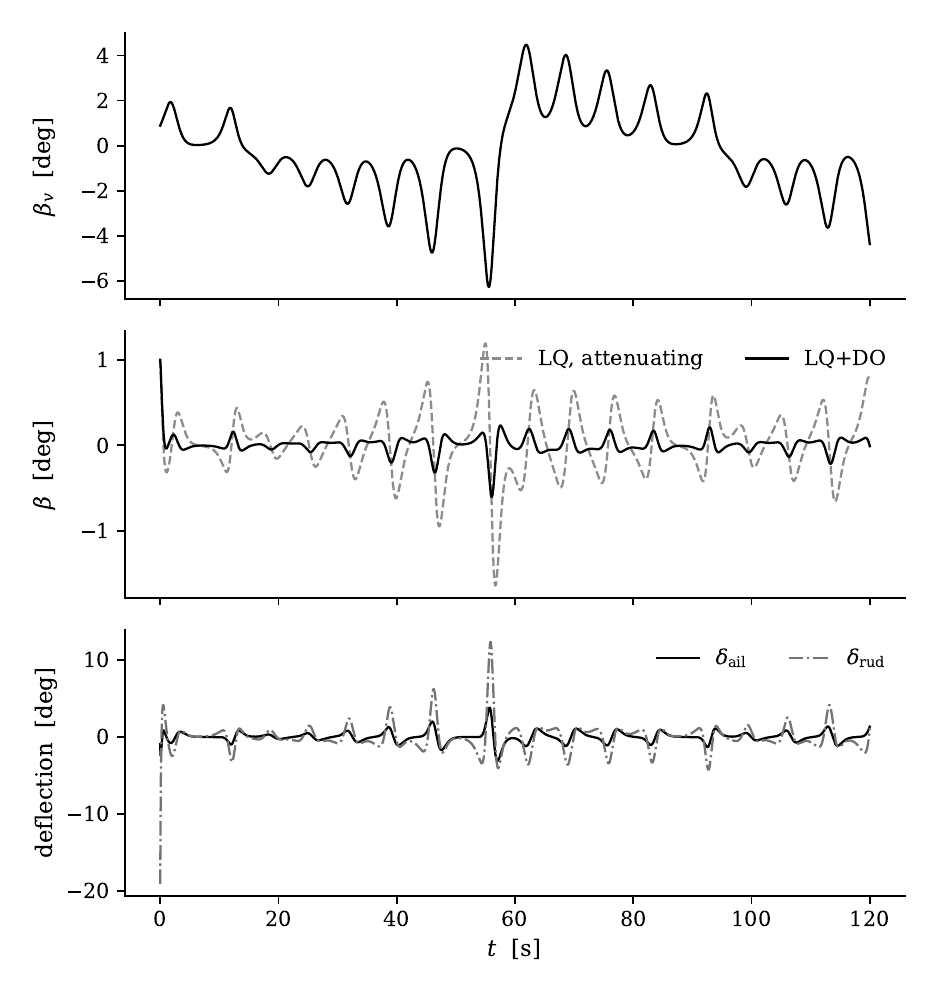}
\caption{Vortex angle; sideslip under LQ attenuation and under LQ+DO (BDO
coincides with LQ+DO at this scale); deflections under LQ+DO.}
\label{fig:sim}
\end{figure}

\subsection{Persistent exogenous vortex: BDO, LQ+DO, full-information limit}\label{sec:persist}

This is Condition~II of Table~\ref{tab:cond}. The exosystem is (\ref{eq:X})--(\ref{eq:Z}) with the airmass origin moving
in one dimension, $\dot y=-\gamma X\sin(\pi\omega y)$,
$\beta_v=-(\gamma/V_T)X\sin(\pi\omega y)$, the one-dimensional form of the
airmass motion, and with $X$, $Y$, $Z$ carrying the classic nondimensional Lorenz magnitudes and
the classic constants $(10,28,1,1,8/3,1)$ scaled by $0.1$~s$^{-1}$:
$\theta_{11}=\theta_{12}=1$, $N_z=1$, $\rho g\theta_{21}=2.8$,
$\theta_{22}=\theta_{23}=0.1$, $\theta_{31}=0.267$, $\theta_{32}=0.1$, all in
s$^{-1}$, so that the vortex rate is of order $1$~s$^{-1}$; $\gamma=1.48$~ft/s
per unit of $X$ converts $X$ to a lateral velocity, and
$\omega=0.00152$~ft$^{-1}$ gives a cell width of $656$~ft. The dimensional
coefficients of Section~\ref{sec:airmass} are used from there on. The gains
are (\ref{eq:Kbdoval}) for backstepping and (\ref{eq:Kval}) for LQ, with
$k=20$ for both; the initial state is $\beta=1^\circ$, $\ps=1^\circ$/s,
$\hat d=0$. The attenuating runs are the same control laws with $\hat d$ set to zero. The
full-information runs replace $\hat d$ by the true $d$ and add the $\dot d$
term, as after (\ref{eq:law}) and (\ref{eq:lqerr}); they are the limit of
output regulation. The results
are in Table~\ref{tab:persist}.

\begin{table}[t]
\caption{Persistent vortex, Condition~II: sideslip rms and peak under six control laws, rms as percentage of the vortex angle, and rudder rms.}
\label{tab:persist}
\centering
\begin{tabular}{lcccc}
\toprule
& rms & peak & rms, \% of vortex & rudder rms\\
\midrule
vortex $\beta_v$        & $2.33^\circ$   & $6.27^\circ$ & $100$ & ---\\
$\beta$, LQ attenuating  & $0.47^\circ$   & $1.64^\circ$ & $20$ & $1.7^\circ$\\
$\beta$, LQ+DO          & $0.117^\circ$  & $0.61^\circ$ & $5.0$ & $2.2^\circ$\\
$\beta$, LQ full information & $0$ & $0$ & $0$ & $1.9^\circ$\\
$\beta$, backstepping attenuating  & $0.66^\circ$   & $2.05^\circ$ & $28$ & $1.4^\circ$\\
$\beta$, BDO          & $0.104^\circ$  & $0.51^\circ$ & $4.5$ & $2.1^\circ$\\
$\beta$, backstepping full information & $0$ & $0$ & $0$ & $1.9^\circ$\\
\bottomrule
\end{tabular}
\end{table}

The LQ+DO and BDO sideslip traces are indistinguishable at the
scale of Figure~\ref{fig:sim}; the two control laws use $\hat d$ identically. For
backstepping the measured ratio BDO/attenuating is $0.158$; applying
(\ref{eq:tfwo}) and (\ref{eq:tfw}) to the spectrum of the simulated $d$ gives
$0.157$, and at the single rate $\omega_L=1.16$~s$^{-1}$, the rms of
$\ddot\beta_v/\dot\beta_v$, (\ref{eq:ratio}) gives $0.17$. LQ attenuation
does better than backstepping attenuation because its poles are faster. The
peak deflections under LQ+DO are $3.8^\circ$ aileron and $19^\circ$ rudder,
the latter in the initial transient. The full-information runs hold sideslip at
zero to solver accuracy at no more rudder than the DO runs: exact
rejection requires no additional control-deflection amplitude in this
ideal-actuator simulation, and everything output
regulation can add over LQ+DO is the $0.117^\circ$. These runs take $d$ and
$\dot d$ from the simulated exosystem state; they are diagnostics, not
implementable controllers. No state observer of the exosystem has been simulated.

\subsection{DO against attenuating feedback}\label{sec:doatt}

The two control laws share the state feedback $-K$ and differ by the rank-one term
$G\,H(s)$ of (\ref{eq:comp}). Table~\ref{tab:trade} collects what the tests
above establish for each.

\begin{table}[t]
\caption{Disturbance-unaware LQ against LQ with the disturbance observer.}
\label{tab:trade}
\centering
\begin{tabular}{>{\raggedright\arraybackslash}p{3.0cm}>{\raggedright\arraybackslash}p{5.2cm}>{\raggedright\arraybackslash}p{5.2cm}}
\toprule
& disturbance-unaware LQ & LQ with disturbance observer\\
\midrule
attenuation at the dutch roll, $4$--$6$~rad/s window, actuators in
& $-2$ to $-4$~dB (Figure~\ref{fig:sweep})
& $-8$ to $-12$~dB; $6$--$8$~dB more, the gain of the DO being (\ref{eq:ratio}), obtained where loop gain is not permitted\\
persistent vortex, Section~\ref{sec:persist} & $20\%$ of the vortex sideslip left & $5\%$ left; the full-information limit is zero\\
ten-second transient, cell vortex, Section~\ref{sec:tensec} & $0.35^\circ$ rms & $0.25^\circ$ rms, ratio $0.70$; less than at the dutch-roll frequency because the circulation spreads the vortex spectrum\\
loop gain at the surfaces & $KP_aP$ & the same, $H(s)P(s)\equiv0$; margins identical for every $k$\\
order & $0$ & $1$, the filter $k/(s+k)$\\
model needed & $A$, $B$ & $A$, $B$ and the side-force row of (\ref{eq:beta}); nothing of the airmass\\
what bounds the gain & margins, structural modes & sensor noise on $\beta$ through the direct term $-kG$; the attenuation saturates at $k$ of two to four times the crossover\\
what it cannot do & \multicolumn{2}{>{\raggedright\arraybackslash}p{10.8cm}}{remove the vortex sideslip: it enters (\ref{eq:beta}) directly, and the residual at the dutch roll is set by the bandwidth}\\
\bottomrule
\end{tabular}
\end{table}

The judgment Table~\ref{tab:trade} supports: the DO is justified whenever the vortex rate
lies inside the loop bandwidth and the sideslip measurement is clean enough
for the direct term; it requires no model of the airmass, and it leaves the
input margins unchanged. Disturbance-unaware feedback is the choice when
sideslip is measured poorly, since then $k$ must be small and the
contribution of the DO is small with it. Neither removes the WR; both
reduce it, by the amounts above.

\section{Output regulation: cost of exact rejection}\label{sec:or}

Exact rejection of the vortex sideslip, in place of the residual of the DO, is
nonlinear output regulation with an internal model of the chaos. This
section builds that regulator to measure its cost: the full-information
control law (\ref{eq:fi}) sets the target, a closed equation
(\ref{eq:closure}) for the vortex rate is the internal model, and two
observers realize it, compared with the DO in Table~\ref{tab:options}. The
setting is the regulated output (\ref{eq:out}) and the exosystem
(\ref{eq:X})--(\ref{eq:Z}) with the one-dimensional airmass motion of
Section~\ref{sec:persist}, $\dot y=-\gamma X\sin(\pi\omega y)$,
$\beta_v=-(\gamma/V_T)X\sin(\pi\omega y)$, autonomous, with $X$, $Y$, $Z$,
$y$ unmeasured; the cell
(\ref{eq:y})--(\ref{eq:bg}) with $p_b=0$ is polynomial in the same way with
two more trigonometric states. We write $a=Y_\beta/V_T$,
$b_r=Y_{\rs}/V_T-1$ and $\varrho=\rho g\,\theta_{21}$.

\noindent\textbf{Full-information control law.} Every control law of this
section is the backstepping control law (\ref{eq:law}) with its first row replaced by
$v-a_\zeta^{\!\top}x$, which turns (\ref{eq:nf}) into $\dot\eta=-c_3\eta$
and $\ddot\beta=v-q$, $q=a\,d+\dot d$: after the roll loop, sideslip is a
double integrator driven by the unmeasured signal $q$ generated by the
exosystem. The control laws differ only in $v$: attenuating backstepping is
$v=-k_v\zeta_2-k_p\beta$ with $k_v=c_1+c_2$, $k_p=1+c_1c_2$, BDO is
(\ref{eq:law}) itself, and the full-information control law is
\begin{equation}
v=a\,d+\dot d-k_p\beta-k_v(\zeta_2-d),
\label{eq:fi}
\end{equation}
under which $\ddot\beta+k_v\dot\beta+k_p\beta=0$: exact rejection, which
requires $d$ and $\dot d$ without lag. The regulator equations are solved in
closed form, zero error requiring $\zeta_2=d$, $\rs=d/b_r$ and
$v=a\,d+\dot d$, and the exosystem enters only through $d$ and $\dot d$; no
Lorenz state appears. What remains is to generate $d$ and $\dot d$ from the measurements
without lag: the internal model.

\noindent\textbf{Closed equation for the vortex rate.} With $S=\sin\pi\omega y$, $C=\cos\pi\omega y$, $\nu=\pi\omega\gamma$,
$\kappa=\gamma/V_T$ and $w=(X,Y,Z,S,C)$, the exosystem
(\ref{eq:X})--(\ref{eq:Z}) with $\dot S=-\nu XSC$, $\dot C=\nu XS^2$ is
polynomial, $\dot w=F(w)$ with $S^2+C^2=1$ invariant, and so are the vortex
rate $d=H_0(w)=\kappa S(\theta_{11}X-\theta_{12}Y+\nu X^2C)$ and its
derivatives $d^{(j)}=H_j(w)$, $H_{j+1}=\nabla H_j\cdot F$. A quadratic function of the Lorenz states gives a compact, forward-invariant
absorbing set $W$, entered in finite time from every initial state, on which
every $H_j$ is bounded. In the polynomial ring in $(w,\bar w)$ the ideals
generated by $H_j(w)-H_j(\bar w)$, $j<N$, form an ascending chain, which
stabilizes; hence for some finite $N$, $H_N(w)-H_N(\bar w)$ is a polynomial
combination of those differences, and with $T=(H_0,\ldots,H_{N-1})$ there
is $L$ with $|H_N(w)-H_N(\bar w)|\le L\,|T(w)-T(\bar w)|$ on $W$. The
extension
\begin{equation}
\Phi(z)={\rm sat}_{[-M,M]}\min_{w\in W}\big\{H_N(w)+L\,|z-T(w)|\big\},
\qquad M>\max_W|H_N|,
\label{eq:Phi}
\end{equation}
is bounded, globally Lipschitz with constant $L$, and satisfies
\begin{equation}
d^{(N)}=\Phi\big(d,\dot d,\ldots,d^{(N-1)}\big)
\label{eq:closure}
\end{equation}
along every trajectory in $W$: the vortex rate obeys an autonomous $N$-th
order equation in which no Lorenz state appears, and that equation is the
internal model. No Lorenz state is reconstructed or needs to be; at a zero
of the vortex profile distinct Lorenz motions produce identical measurements,
and there is nothing to reject. For the exosystem of Section~\ref{sec:persist},
the numerical tests of Appendix~\ref{app:closure} indicate $N=5$ on the
tested attractor region, with $L$ up to $8.4$ on the attractor
in units where each $d^{(j)}$ is scaled by its rms; (\ref{eq:Phi}) is a
global minimization over $W$ at each evaluation, with no efficient exact
realization known.

\noindent\textbf{Regulator with a high-gain observer.} The coordinates $r_0=-\beta$,
$r_i=d^{(i-1)}$, $i=1,\ldots,N$, obey $\dot r_0=r_1-\zeta_2$,
$\dot r_i=r_{i+1}$ for $i<N$, $\dot r_N=\Phi(r_1,\ldots,r_N)$, with only
$r_0$ measured and $\zeta_2$ a known input. With $m=N+1$,
$h_i=\binom{m}{i+1}$, $\ell\ge1$ and $\epsilon_o=-\beta-\hat r_0$,
\begin{subequations}\label{eq:hgo}
\begin{eqnarray}
\dot{\hat r}_0&=&\hat r_1-\zeta_2+\ell\,h_0\,\epsilon_o,\qquad
\dot{\hat r}_i=\hat r_{i+1}+\ell^{i+1}h_i\,\epsilon_o\ \ (1\le i<N), \label{eq:hgo0}\\
\dot{\hat r}_N&=&\Phi(\hat r_1,\ldots,\hat r_N)+\ell^{N+1}h_N\,\epsilon_o \label{eq:hgoi}\\
v&=&a\,\hat r_1+\hat r_2-k_p\beta-k_v(\zeta_2-\hat r_1) \label{eq:vor}
\end{eqnarray}
\end{subequations}
is the regulator, with (\ref{eq:law}) as above and the offline map (\ref{eq:Phi}); no
measured signal is differentiated. With $E_i=\hat r_i-r_i$,
$\varepsilon_i=E_i/\ell^i$, $H=A_m-he_1^{\!\top}$ ($A_m$ the shift,
characteristic polynomial $(s+1)^m$), and $P=P^{\!\top}>0$ solving
$H^{\!\top}P+PH=-I$,
\begin{equation}
\dfrac{d}{dt}\big(\varepsilon^{\!\top}P\varepsilon\big)\le
-\big(\ell-2L\,|Pe_m|\big)|\varepsilon|^2 ,\qquad
\ddot\beta+k_v\dot\beta+k_p\beta=(a+k_v)E_1+E_2 ,
\label{eq:conv}
\end{equation}
so for $\ell>2L|Pe_m|$ the estimation error and then $\beta$ decay
exponentially, uniformly over exosystem trajectories in $W$; with exact
$\Phi$, plant and measurement. This is a high-gain observer with injection
up to $\ell^{N+1}$. For $N=5$, $|Pe_m|=114$, and (\ref{eq:conv}) requires
$\ell>1.9\times10^3$ for $L=8.4$ and $\ell>230$ even for $L=1$; the
injection gains reach $\ell^6\ge10^{14}$, and with the guaranteed gain the
regulator is not implementable on a vane signal. Nor is the realization
exact: with an implemented map $\Phi+\eta_\Phi$, $|\eta_\Phi|\le\delta_\Phi$
on the operating region, the residual in $(\beta,\dot\beta)$ is bounded by a
multiple of $\ell^{2-N}\delta_\Phi$, proportional to the uniform fitting
error, and a small error along one trajectory does not supply
$\delta_\Phi$.

\noindent\textbf{Regulator with a KKL observer.} The high-gain
observer (\ref{eq:hgo}) is the DO (\ref{eq:dobtf}) carried to sixth order with $\Phi$ as
the top equation; the exosystem enters only through the table $\Phi$, and
the single injection point forces $\ell^{6}$. An observer that uses the
polynomial exosystem itself, instead of the closure $\Phi$, is a Kazantzis--Kravaris/Luenberger (KKL)
observer \cite{AP06,MPI07}: with $A\in\mathbb{R}^{m\times m}$ Hurwitz,
$B\in\mathbb{R}^m$, and $r_0=-\beta$,
\begin{equation}
\dot z=Az-B\beta+A^{-1}B\,\zeta_2,\qquad
\chi=z-A^{-1}B\beta,\qquad
v=\rho(\chi)-k_p\beta-k_v\big(\zeta_2-\rho_0(\chi)\big),
\label{eq:kkl}
\end{equation}
where $z$ converges, at the eigenvalues of $A$ and for every input
$\zeta_2$, to $\tau(w)-A^{-1}B\,r_0$, so that $\chi$ converges to
$\tau(w)$, with $\tau$ the solution of
$\nabla\tau\cdot F=A\tau+A^{-1}B\,H_0$, which exists when those eigenvalues
are more negative than the backward growth rate of the exosystem, of the
order of $\theta_{11}+\theta_{22}+\theta_{31}$: order one, not a power of a
gain. The readouts $\rho$, $\rho_0$, with
\begin{equation}
\rho(\tau(w))=q(w),\qquad \rho_0(\tau(w))=H_0(w),
\label{eq:readout}
\end{equation}
return $q=aH_0+H_1$ and $H_0$ from $\chi$;
they exist for generic $A$ with $m\ge9$ when the exosystem is
backward-distinguishable from $H_0$ \cite{AP06}, and are built offline by
running (\ref{eq:kkl}) on a long attractor trajectory of the known exosystem
and fitting. Against the high-gain observer: order nine or ten instead of six, and a
readout fitted rather than constructed, with an error bound not derived
here. For it: gains of order one against $\ell^6\ge10^{14}$, and noise
entering linearly through $B$. The broadband spectrum and the positive
Lyapunov exponent of the Lorenz block are not obstructions: (\ref{eq:kkl})
is driven by the measured $\beta$ and synchronizes, it does not forecast.
The closure (\ref{eq:closure}) is proved directly, without the recurrence
and immersion assumptions of \cite{BI04}; \cite{MP08} is the reference for
practical regulation with an approximate $\rho$; adaptive regulators for
linear exosystems with unknown frequencies \cite{MT11} do not apply. With
the parameters unknown, nothing here applies: (\ref{eq:closure}),
(\ref{eq:Phi}) and (\ref{eq:kkl}) depend on all nine.

\begin{table}[t]
\caption{Disturbance observer against the two output-regulation routes.}
\label{tab:options}
\centering
\begin{tabular}{>{\raggedright\arraybackslash}p{2.9cm}>{\raggedright\arraybackslash}p{3.7cm}>{\raggedright\arraybackslash}p{3.6cm}>{\raggedright\arraybackslash}p{3.8cm}}
\toprule
& DO with LQ or backstepping, Section~3 & KKL route & high-gain observer route\\
\midrule
exosystem model & none & needed & needed, polynomial form\\
parameters & none & all nine & all nine\\
order & one filter state & $9$ to $10$ & $6$ for the tested $N=5$ closure\\
nominal rejection & residual, (\ref{eq:tfw}) & exact, given $\rho$ & exact, (\ref{eq:conv})\\
gain & bandwidth $k$; input margins those of LQR & eigenvalues of $A$, order one & $\ell^6\ge10^{14}$ by (\ref{eq:conv})\\
uses Lorenz structure & no & yes, through $\tau$ & only through the table $\Phi$\\
\bottomrule
\end{tabular}
\end{table}

\paragraph*{What exact regulation adds over the DO.}
Table~\ref{tab:options} compares the three. What output regulation can add
over the DO is known exactly: (\ref{eq:dobtf}) is the high-gain observer
(\ref{eq:hgo}) truncated at $N=1$ with $\dot r_1=0$, residual driven by
$\dot d$, while closure with $\Phi$ leaves none, so the advantage is the DO
residual itself. In the simulation of Section~\ref{sec:persist} attenuation leaves
$20\%$ of the vortex angle in sideslip under LQ and $28\%$ under backstepping;
the DO leaves $5.0\%$ and $4.5\%$, and that remainder, $0.12^\circ$ rms, is
all that output regulation can remove, at no additional control-deflection amplitude in this ideal-actuator
simulation;
a sideslip vane resolves a few tenths of a degree. The high-gain observer is
ruled out by its gain; the KKL route, gains of order one, is the one to
build if any is built, and it is justified only if that residual matters
at the resolution of the vane and under noise. Neither has been built: the
full-information limit (\ref{eq:fi}) has been simulated with $d$, $\dot d$
taken from the exosystem state, and the closure order and Lipschitz constant
computed on the attractor. With the parameters unknown, nothing beyond the
DO is available.

\section{Conclusions}

In the proposed model, WR is defined as an eight-state dynamical system, not a two-state roll oscillator with a
nonlinear damping. Three states are the aircraft sideslip and the roll and
yaw rates, and they are governed linearly; the aircraft dynamics at the WR
condition are linear and lightly stable. Additional five states represent the
airmass vorticial dynamics, and they are governed bilinearly: the three Lorenz states of the convection that the
lift reaction drives, with the products $XZ$ and $XY$, and the two
coordinates of the airmass origin, with the product of $X$ and the cell
velocity field and the cross product of the body rotation with position
that the non-inertial body frame introduces. When the aircraft vertical
load factor is sufficiently large, the convection becomes chaotic and its
primary natural frequency of the nonzero equilibrium falls near the
dutch-roll frequency. As a result, the linear, lightly damped aircraft
dynamics respond at resonance, and the eight states emulate the WR
phenomenon. From a controls perspective, WR is thereby a disturbance
rejection problem, with achievable attenuation levels defined by the model
itself: the vortex sideslip enters the sideslip equation (\ref{eq:beta}), where the
contribution of the control surfaces is minimal, so the WR oscillations can
be attenuated, but not entirely removed.

The control design proposed in this paper is focused on attenuation of the WR
induced transients. Each part of the model is handled according to what is
known about it. The aircraft model is linear and known, so the aircraft is
controlled by LQ feedback, which comes with guaranteed margins. The airmass
is chaotic and its states are not measured, so the vortex sideslip is
estimated by a one-state DO that uses no airmass model and assumes only
that the vortex sideslip is differentiable. The DO leaves the loop gain at
the surfaces unchanged, and it attenuates the system resonance by a further
$6$--$8$~dB at frequencies where the loop gain cannot be raised. The exact alternative, nonlinear output
regulation with an internal model of the chaos, requires all nine airmass
parameters and an order-six-to-ten regulator, and improves performance of the observer
by $5\%$ of the vortex angle, below the resolution of a vane. A one-state
addition to a standard lateral-directional control law recovers nearly all of what
the most elaborate regulator could.

\appendix
\section{Conditions, data and computations}

\subsection{F-16 data}\label{app:data}

\noindent\textbf{Condition~II of Table~\ref{tab:cond},
$\alpha=30^\circ$, $V_T=230$~ft/s, sea level.} The S\&C derivatives are $L_\beta=-10.2$,
$L_{\ps}=-0.89$, $L_{\rs}=2.64$, $N_\beta=0.41$, $N_{\ps}=0.076$,
$N_{\rs}=-0.35$, $Y_\beta/V_T=-0.15$, $Y_{\ps}=Y_{\rs}=0$,
$L_{\delta_\ail}=-5.3$, $N_{\delta_\ail}=0.18$, $L_{\delta_\rud}=1.47$,
$N_{\delta_\rud}=-0.84$, in s$^{-1}$ and s$^{-2}$. The open-loop eigenvalues
of $A$ are $-0.853\pm0.72j$ and $+0.316$, a divergence with a three-second time
constant. Single-surface controllability is in Table~\ref{tab:single}.

\begin{table}[t]
\caption{Single-surface controllability, first data set: determinant and condition number.}
\label{tab:single}
\centering
\begin{tabular}{lcc}
\toprule
& $\det[\,b\ Ab\ A^2b\,]$ & $\kappa$\\
\midrule
aileron & $0.43$ & $81$\\
rudder  & $4.4$  & $8.4$\\
\bottomrule
\end{tabular}
\end{table}

\noindent\textbf{Condition~I of Table~\ref{tab:cond}, $\alpha=30^\circ$,
$50$~psf.} The lateral-directional model is from
the NASA F-16 aerodynamic tables \cite{NOG79} as coded in \cite{SL}, at
$\alpha=30^\circ$, $\beta=0$, $\bar q=50$~psf, sea level, $V_T=205$~ft/s;
$C_{l_\beta}$, $C_{n_\beta}$ from the $\beta=5^\circ$ column, rotary
derivatives and control derivatives at $\beta=0$, $m=637$~slug,
$S=300$~ft$^2$, $b=30$~ft, $I_x=9496$, $I_z=63100$,
$I_{xz}=982$~slug\,ft$^2$; states $(\beta,\ps,\rs)$ in rad and rad/s,
surfaces in rad:
\begin{equation}
A=\begin{pmatrix}-0.132&0.007&-0.998\\-6.94&0.370&1.96\\4.24&-0.316&-1.44\end{pmatrix},\qquad
B=\begin{pmatrix}0.0069&0.0189\\-3.60&0.64\\2.17&-1.12\end{pmatrix}.
\label{eq:highalpha}
\end{equation}
The first row of $B$, the effectiveness of the control surfaces on the side
force, is dropped in the
designs except that of the servo of Section~\ref{sec:servo}, and from the plant and the DO in all simulations except those of
Section~\ref{sec:trackff}; retained in the plant and in the DO
(\ref{eq:dobtf}), it changes the sideslip and roll-rate figures of the two
control laws of Figure~\ref{fig:track} by at most $6\%$.
The eigenvalues are $-0.524\pm2.02j$ and $-0.149$; with the rotary damping
derivatives $C_{Y_p}$, $C_{Y_r}$, $C_{l_p}$, $C_{l_r}$, $C_{n_p}$, $C_{n_r}$
multiplied by $0.2$ they are $-0.157\pm2.06j$, $-0.030$. The gains of
Section~\ref{sec:tensec} are
\begin{equation}
K=\begin{pmatrix}-15.7&-0.32&2.72\\8.45&-0.33&-2.05\end{pmatrix}.
\label{eq:Khigh}
\end{equation}

\subsection{Airmass origin}\label{app:origin}

$(y,z)$ is the position of the airmass origin relative to the body-fixed
frame. With $p_b=0$, (\ref{eq:y})--(\ref{eq:z}) is
$\dot y=-\partial_z\psi$, $\dot z=\partial_y\psi$ with the streamfunction
$\psi=(\gamma_y/\pi\omega_z)\,X\sin(\pi\omega_yy)\sin(\pi\omega_zz)$, so
the origin moves along a level curve of $\sin(\pi\omega_yy)\sin(\pi\omega_zz)$,
a closed curve inside the cell, forward while $X>0$ and back while $X<0$;
the speed is set by $X$, the path is not. Along the path $v_v$ vanishes only
on the lines $y=j/\omega_y$ and $z=j/\omega_z$, which the origin crosses,
so the vortex sideslip persists. The exceptions are the level curve zero, the cell
boundary, on which $y=0$ (say) is invariant and $v_v\equiv0$, and the cell
center $(1/2\omega_y,1/2\omega_z)$, a rest point; a start on the boundary,
such as $(0,-0.1)$, produces no vortex sideslip until $p_b$ moves the origin off it.
The exact form of (\ref{eq:bg}) requires $\gamma_yX_{\max}<V_T$, with
$X_{\max}$ the extent in $X$ of the absorbing set of
(\ref{eq:X})--(\ref{eq:Z}).

\subsection{Closure order}\label{app:closure}

For the exosystem of Section~\ref{sec:persist}, $H_0,\ldots,H_8$ were generated
symbolically from the polynomial exosystem of Section~\ref{sec:or}, polynomials of degree
$4+2j$ with $358$ terms at $j=8$, evaluated on a $2800$~s attractor
trajectory of $1.4\times10^5$ points, and scaled by their rms there. The rank
of $dT_N$ on the exosystem manifold is four, its dimension, from $N=4$ on.
For eight attractor targets, every solution of $T_N(w)=T_N(w_\ast)$ in the
box $|X|\le25$, $|Y|\le35$, $0\le Z\le50$ was sought from $300$
Levenberg--Marquardt starts: $N=4$ gives two to eight preimages per target,
each with $H_4$ different from $H_4(w_\ast)$ by up to $0.32$; $N=5$ and
$N=6$ give the target only. Over nearest-neighbour pairs in $T_N$-space,
pairs adjacent in time excluded, the quotient
$|H_N(w)-H_N(\bar w)|/|T_N(w)-T_N(\bar w)|$ has maximum $2.8$ at $N=5$,
and the local constant $\|\nabla H_N\,(dT_N)^{+}\|$ has median $0.8$ and
maximum $8.4$. For $N=5$, $m=6$, $h_i=\binom{6}{i+1}$, $|Pe_m|=114$ in
(\ref{eq:conv}). These establish $N=5$ numerically on an attractor-sized
box; they do not cover $W$ and do not construct $\Phi$.

\section*{Acknowledgment}
The problem formulations, ideas, writing style, design and analysis results reported in this paper are
original and belong to the authors. Simulation test cases were developed
with the assistance of Claude and ChatGPT under the authors' correction and
complete verification.

\addcontentsline{toc}{section}{References}
\bibliographystyle{unsrt}
\bibliography{refs-v150}

@book{LW, author={Lavretsky, Eugene and Wise, Kevin A.}, title={Robust and Adaptive Control with Aerospace Applications}, edition={2nd}, publisher={Springer}, year={2024}, note={doi:\url{10.1007/978-3-031-38314-4}}}

@article{AP06, author={Andrieu, Vincent and Praly, Laurent}, title={On the existence of a {K}azantzis--{K}ravaris/{L}uenberger observer}, journal={SIAM Journal on Control and Optimization}, volume={45}, number={2}, pages={432--456}, year={2006}, note={doi:\url{10.1137/040617066}}}

@article{MPI07, author={Marconi, Lorenzo and Praly, Laurent and Isidori, Alberto}, title={Output stabilization via nonlinear {L}uenberger observers}, journal={SIAM Journal on Control and Optimization}, volume={45}, number={6}, pages={2277--2298}, year={2007}, note={doi:\url{10.1137/050642344}}}

@article{BI04, author={Byrnes, Christopher I. and Isidori, Alberto}, title={Nonlinear internal models for output regulation}, journal={IEEE Transactions on Automatic Control}, volume={49}, number={12}, pages={2244--2247}, year={2004}, note={doi:\url{10.1109/TAC.2004.838492}}}

@article{MP08, author={Marconi, Lorenzo and Praly, Laurent}, title={Uniform practical nonlinear output regulation}, journal={IEEE Transactions on Automatic Control}, volume={53}, number={5}, pages={1184--1202}, year={2008}, note={doi:\url{10.1109/TAC.2008.923674}}}

@techreport{NOG79, author={Nguyen, L. T. and Ogburn, M. E. and Gilbert, W. P. and Kibler, K. S. and Brown, P. W. and Deal, P. L.}, title={Simulator study of stall/post-stall characteristics of a fighter airplane with relaxed longitudinal static stability}, institution={NASA}, number={TP-1538}, year={1979}}

@book{SL, author={Stevens, Brian L. and Lewis, Frank L. and Johnson, Eric N.}, title={Aircraft Control and Simulation}, edition={3rd}, publisher={Wiley}, year={2015}, note={doi:\url{10.1002/9781119174882}}}

@techreport{CG77, author={Chambers, J. R. and Grafton, S. B.}, title={Aerodynamic characteristics of airplanes at high angles of attack}, institution={NASA}, number={TM-74097}, year={1977}}

@inproceedings{NWB85, author={Nguyen, L. T. and Whipple, R. D. and Brandon, J. M.}, title={Recent experiences of unsteady aerodynamic effects on aircraft flight dynamics at high angle of attack}, booktitle={Unsteady Aerodynamics---Fundamentals and Applications to Aircraft Dynamics, AGARD-CP-386}, address={G\"ottingen}, year={1985}}

@techreport{NGO80, author={Nguyen, Luat T. and Gilbert, William P. and Ogburn, Marilyn E.}, title={Control-system techniques for improved departure/spin resistance for fighter aircraft}, institution={NASA}, number={TP-1689}, year={1980}}

@article{Ericsson84, author={Ericsson, Lars E.}, title={The fluid mechanics of slender wing rock}, journal={Journal of Aircraft}, volume={21}, number={5}, pages={322--328}, year={1984}, note={doi:\url{10.2514/3.44967}}}

@article{HL85, author={Hsu, C.-H. and Lan, C. Edward}, title={Theory of wing rock}, journal={Journal of Aircraft}, volume={22}, number={10}, pages={920--924}, year={1985}, note={doi:\url{10.2514/3.45225}}}

@article{EMN89, author={Elzebda, Jamal M. and Nayfeh, Ali H. and Mook, Dean T.}, title={Development of an analytical model of wing rock for slender delta wings}, journal={Journal of Aircraft}, volume={26}, number={8}, pages={737--743}, year={1989}, note={doi:\url{10.2514/3.45833}}}

@article{AN94, author={Arena, Andrew S. and Nelson, Robert C.}, title={Experimental investigations on limit cycle wing rock of slender wings}, journal={Journal of Aircraft}, volume={31}, number={5}, pages={1148--1155}, year={1994}, note={doi:\url{10.2514/3.46625}}}

@article{Katz99, author={Katz, Joseph}, title={Wing/vortex interactions and wing rock}, journal={Progress in Aerospace Sciences}, volume={35}, number={7}, pages={727--750}, year={1999}, note={doi:\url{10.1016/S0376-0421(99)00004-4}}}

@article{NP03, author={Nelson, Robert C. and Pelletier, Alain}, title={The unsteady aerodynamics of slender wings and aircraft undergoing large amplitude maneuvers}, journal={Progress in Aerospace Sciences}, volume={39}, number={2--3}, pages={185--248}, year={2003}, note={doi:\url{10.1016/S0376-0421(02)00088-X}}}

@article{GR02, author={Go, Tiauw H. and Ramnath, Rudrapatna V.}, title={Analysis of the two-degree-of-freedom wing rock in advanced aircraft}, journal={Journal of Guidance, Control, and Dynamics}, volume={25}, number={2}, pages={324--333}, year={2002}, note={doi:\url{10.2514/2.4885}}}

@article{GK94, author={Goman, Mikhail and Khrabrov, Alexander}, title={State-space representation of aerodynamic characteristics of an aircraft at high angles of attack}, journal={Journal of Aircraft}, volume={31}, number={5}, pages={1109--1115}, year={1994}, note={doi:\url{10.2514/3.46618}}}

@inproceedings{LD97, author={Liebst, Bradley S. and DeWitt, Bruce R.}, title={Wing rock suppression in the {F}-15 aircraft}, booktitle={AIAA Atmospheric Flight Mechanics Conference}, note={AIAA Paper 97-3719}, year={1997}}

@mastersthesis{W92, author={Wright, Stephen R.}, title={A wing rock model for the {F}-14{A} aircraft}, school={Naval Postgraduate School}, address={Monterey, CA}, type={Engineer's thesis}, year={1992}}

@phdthesis{S00, author={Saad, Ashraf A.}, title={Simulation and analysis of wing rock physics for a generic fighter model with three degrees-of-freedom}, school={Air Force Institute of Technology}, note={AFIT/DS/ENY/00-01}, year={2000}}

@article{LL93, author={Luo, Jiang and Lan, C. Edward}, title={Control of wing-rock motion of slender delta wings}, journal={Journal of Guidance, Control, and Dynamics}, volume={16}, number={2}, pages={225--231}, year={1993}, note={doi:\url{10.2514/3.20993}}}

@article{SYW95, author={Singh, Sahjendra N. and Yim, Woosoon and Wells, William R.}, title={Direct adaptive and neural control of wing-rock motion of slender delta wings}, journal={Journal of Guidance, Control, and Dynamics}, volume={18}, number={1}, pages={25--30}, year={1995}, note={doi:\url{10.2514/3.56652}}}

@article{MK96, author={Monahemi, Mahmood M. and Krstic, Miroslav}, title={Control of wing rock motion using adaptive feedback linearization}, journal={Journal of Guidance, Control, and Dynamics}, volume={19}, number={4}, pages={905--912}, year={1996}, note={doi:\url{10.2514/3.21717}}}

@article{WCGW17, author={Wu, Dawei and Chen, Mou and Gong, Huajun and Wu, Qingxian}, title={Robust backstepping control of wing rock using disturbance observer}, journal={Applied Sciences}, volume={7}, number={3}, pages={219}, year={2017}, note={doi:\url{10.3390/app7030219}}}

@article{LTWI22, author={Li, Dongyang and Tsourdos, Antonios and Wang, Zhenhua and Ignatyev, Dmitry}, title={Nonlinear analysis for wing-rock system with adaptive control}, journal={Journal of Guidance, Control, and Dynamics}, volume={45}, number={11}, pages={2174--2181}, year={2022}, note={doi:\url{10.2514/1.G006775}}}

@article{LCZSQ18, author={Lyu, Yongfeng and Cao, Yunyun and Zhang, Weiguo and Shi, Jingping and Qu, Xiaobo}, title={Dynamic surface control design of post-stall maneuver under unsteady aerodynamics}, journal={Aerospace Science and Technology}, volume={80}, pages={269--280}, year={2018}, note={doi:\url{10.1016/j.ast.2018.06.037}}}

@article{SJL20, author={Sedky, Girguis and Jones, Anya R. and Lagor, Francis D.}, title={Lift regulation during transverse gust encounters using a modified {G}oman--{K}hrabrov model}, journal={AIAA Journal}, volume={58}, number={9}, pages={3788--3798}, year={2020}, note={doi:\url{10.2514/1.J059127}}}

@inproceedings{ML26, author={Menner, Marcel and Lavretsky, Eugene}, title={Modeling unsteady aircraft aerodynamics using {L}orenz attractor: A reduced-order approach for wing rock}, booktitle={AIAA SciTech 2026 Forum}, note={AIAA 2026-1171; arXiv:2511.07349, doi:\url{10.2514/6.2026-1171}}, year={2026}}

@article{Saltzman62, author={Saltzman, Barry}, title={Finite amplitude free convection as an initial value problem---{I}}, journal={Journal of the Atmospheric Sciences}, volume={19}, number={4}, pages={329--341}, year={1962}, note={doi:\url{10.1175/1520-0469(1962)019<0329:FAFCAA>2.0.CO;2}}}

@article{Lorenz63, author={Lorenz, Edward N.}, title={Deterministic nonperiodic flow}, journal={Journal of the Atmospheric Sciences}, volume={20}, number={2}, pages={130--141}, year={1963}, note={doi:\url{10.1175/1520-0469(1963)020<0130:DNF>2.0.CO;2}}}

@article{MT11, author={Marino, Riccardo and Tomei, Patrizio}, title={An adaptive learning regulator for uncertain minimum phase systems with undermodeled unknown exosystems}, journal={Automatica}, volume={47}, number={4}, pages={739--747}, year={2011}, note={doi:\url{10.1016/j.automatica.2011.01.019}}}

@article{Ericsson95, author={Ericsson, Lars E.}, title={Wing rock analysis of slender delta wings, review and extension}, journal={Journal of Aircraft}, volume={32}, number={6}, pages={1221--1226}, year={1995}, note={doi:\url{10.2514/3.46867}}}

@article{Mabey97, author={Mabey, Dennis G.}, title={Similitude relations for buffet and wing rock on delta wings}, journal={Progress in Aerospace Sciences}, volume={33}, number={7--8}, pages={481--511}, year={1997}, note={doi:\url{10.1016/S0376-0421(97)00003-1}}}

@article{GGV05, author={Gursul, Ismet and Gordnier, Raymond and Visbal, Miguel}, title={Unsteady aerodynamics of nonslender delta wings}, journal={Progress in Aerospace Sciences}, volume={41}, number={7}, pages={515--557}, year={2005}, note={doi:\url{10.1016/j.paerosci.2005.09.002}}}

@article{LFW23, author={Li, Xiao and Feng, Li-Hao and Wang, Qi-Ming}, title={Wing rock mode and its mechanism of a flying-wing aircraft}, journal={Flow}, volume={3}, pages={E38}, year={2023}, note={doi:\url{10.1017/flo.2023.30}}}

@mastersthesis{Nolan92, author={Nolan, Robert C.}, title={Wing rock prediction method for a high performance fighter aircraft}, school={Air Force Institute of Technology}, address={Wright-Patterson AFB, OH}, type={M.S. thesis}, note={AFIT/GAE/ENY/92J-02, DTIC AD-A256438}, year={1992}}

\end{document}